\documentclass[prx,showpacs,twocolumn,superscriptaddress,longbibliography,floatfix,notitlepage,nofootinbib]{revtex4-2}
\usepackage[utf8]{inputenc}
\usepackage[english]{babel}

\usepackage{natbib}

\usepackage{times}
\usepackage{amsmath}
\usepackage{amsfonts}
\usepackage{amssymb}
\usepackage{dsfont}
\usepackage{graphicx}
\usepackage{physics}
\usepackage{braket}
\usepackage{algorithm}
\usepackage{algpseudocode}
\usepackage[labelformat=simple]{subcaption}
\usepackage{todonotes}
\usepackage{hyperref}
\usepackage{csquotes}
\usepackage[normalem]{ulem}

\newcommand{\e}{\mathrm{e}} 

\newcommand{\figref}[1]{Fig.~\ref{#1}}
\newcommand{\1}{\mathbf{1}}

\begin{document}
	\title{Fault-tolerant error correction for a universal non-Abelian topological quantum computer at finite temperature}
	
	\author{Alexis Schotte}
	\email{alexis.schotte@posteo.net}
	\affiliation{IBM Quantum, IBM Almaden Research Center, San Jose, CA 95120, USA}
	
	\author{Lander Burgelman}
	\affiliation{Department of Physics and Astronomy, Ghent University, Krijgslaan 281, 9000 Gent, Belgium}
	
	\author{Guanyu Zhu}
 \email{guanyu.zhu@ibm.com}
	\affiliation{IBM Quantum, IBM Almaden Research Center, San Jose, CA 95120, USA}
 	\affiliation{IBM T. J. Watson Research Center, Yorktown Heights, NY 10598, USA}

	\begin{abstract}
		We study fault-tolerant error correction in a quantum memory constructed as a two-dimensional model of Fibonacci anyons on a torus, in the presence of thermal noise represented by pair-creation processes and measurement errors. The correction procedure is based on the cellular automaton decoders originating in the works of Gács \cite{gacs1986reliable}
        and Harrington \cite{harrington}.
        Through numerical simulations, we observe that this code behaves fault-tolerantly and that threshold behavior is likely present. Hence, we provide strong evidence for the existence of a fault-tolerant universal non-Abelian topological quantum computer. 
	\end{abstract}
	
	\maketitle

	\section{Introduction}

    Anyons are emergent quasi-particles that exist in two-dimensional condensed matter systems and whose exchange statistics generalize that of Bosons and Fermions. These particles have spurred much interest due to their potential applications for quantum computation. 
    In particular, it was found that with certain types of non-Abelian anyons, a universal quantum computation can be performed by braiding and fusing these particles \cite{freedman2002modular, freedman2002simulation, kitaev2003fault}. 
    An intriguing benefit of this paradigm is that, due to their topological nature, computations are intrinsically robust to perturbations at zero temperature. 
    At non-zero temperature, however, thermal anyonic excitations can corrupt the computation by performing non-trivial braids with the computational anyons. 
    Since systems exhibiting anyonic excitations have a spectral gap $\Delta$, this source of errors can be suppressed to some extent at temperatures $T \ll \Delta/k_B$ as the density of thermal anyons scales as $e^{-\Delta/k_B T}$.
    Alas, this passive protection does not suffice, because the presence of thermal anyons is unavoidable at non-zero temperatures when scaling up the size of computations. 
    Therefore, proficient active error correction schemes for non-Abelian models are paramount for the realization of topological quantum computers. 

    Besides their envisaged use for topological quantum computation, topologically ordered systems (i.e., those that support anyonic excitations on top of their ground space) are also of much interest for quantum error correction. 
    In particular, one of the characteristics of such systems is a robust ground space degeneracy, which allows one to use their ground space as the code space of an error correcting code. 
    This realization led to the discovery of topological quantum error correcting codes, which encode logical quantum states in topologically ordered states of a system of qudits (typically arranged on a two-dimensional lattice). 
    Since their discovery in the 90s, most research has focused exclusively on Abelian topological codes such as the surface code and the color code \cite{bravyi1998quantum, dennis2002topological, kitaev2003fault, wang2009threshold, fowler2012surface, wootton2012high, anwar2014fast, bravyi2014efficient, wootton2015simple, andrist2015error}, which admit an elegant characterization in terms of the stabilizer formalism \cite{gottesman1997stabilizer}.
    Due to their geometrical locality and high error thresholds, these codes are considered to be promising candidates for protecting quantum information from noise in error-corrected quantum computers.
    One of the drawbacks of Abelian topological codes, however, is that they do not allow one to execute a universal set of logical gates in a protected fashion in two dimensions. Hence, they must be supplemented with additional protocols such as magic state distillation \cite{bravyi2005universal} or code switching to higher-dimensional codes \cite{kubica2015universal, bombin2016dimensional} , which introduce a large space-time overhead \cite{campbell2017roads}. 
    Alternatively, there exist non-Abelian topological codes which do not suffer from this inherent limitation, and are able to perform a universal gate set natively within their code space in two dimensions \cite{freedman2002modular}. The trade-off is that such codes go beyond the stabilizer formalism and are therefore very hard to simulate classically.

    While active error correction in Abelian anyon models and Abelian topological codes has been studied extensively, quantum error correction based on non-Abelian anyon models has not enjoyed the same focus. Nevertheless, important progress has been made over the last decade, including both analytical proofs and numerical demonstrations of threshold behavior for various non-Abelian topological error correcting codes \cite{hutter2016continuous, wootton2014error, brell2014thermalization, burton2017classical, schotte2022quantum}. Moreover, syndrome extraction circuits for such non-Abelian string-net codes have been developed in recent years \cite{Bonesteel:2012fl, schotte2022quantum}. In addition, state preparation for non-Abelian codes based on the Kitaev quantum double models via measurements has also been proposed recently for the experimental implementation on qubit lattices \cite{verresen2021efficiently, tantivasadakarn2022shortest}, although further development is still needed in the context of fault-tolerant state preparation.  Notably,  previous studies in this field already include codes based on the Fibonacci anyon model, which is universal for quantum computation \cite{burton2017classical, schotte2022quantum}. In particular, a quantum memory of qubits supporting doubled Fibonacci anyonic excitations was found to have a threshold that lies remarkably close to that of the surface code under similar assumptions \cite{schotte2022quantum}.

    These results, however, all assume perfect syndrome measurements, which are topological charge measurements in this context. As we aim to model more realistic scenarios, we must take faulty measurements into consideration. 
    Again, much is known in the case of Abelian topological codes \cite{harrington, dennis2002topological, raussendorf2007fault, fowler2009high, watson2015fast, herold2017cellular}. 
    For their non-Abelian counterparts, one key result stands out: in Ref.~\cite{dauphinais2017fault} a proof was formulated that topological codes based on non-cyclic anyon models admit a error correction thresholds with faulty topological charge measurements.
    While this result is remarkable, non-cyclic anyon models are not universal for quantum computation, and it remains an open question whether similar claims can be made for universal models.

    In this work, we take a step towards demonstrating that fault-tolerance is indeed possible for universal non-Abelian topological codes. To this end, we define a quantum memory constructed as a two-dimensional model of Fibonacci anyons on a torus. We study active continuous quantum error correction on this model in the presence of thermal noise represented by pair-creation processes, and with faulty syndrome measurements. 
	The correction procedure is based on the cellular automaton decoders originating in the works of Gács \cite{gacs1986reliable} and Harrington \cite{harrington}, and further studied in the context of non-Abelian models in Ref.~\cite{dauphinais2017fault}. 
	Through numerical simulations, we study how the average memory lifetime changes with the error rate. The results indicate that this code is indeed fault-tolerant, which is strong evidence for the existence of fault-tolerant universal non-Abelian codes.

	The structure of this work is as follows.
	In Sec.~\ref{sec:code} we introduce the topological Fibonacci code. We then describe the details of the noise model in Sec.~\ref{sec:noise} and introduce the cellular automaton decoder in Sec.~\ref{sec:decoder}. 
	We proceed by giving an outline of the numerical simulations performed in this work in Sec.~\ref{sec:simulation}.
	Finally, we present the corresponding numerical results in Sec.~\ref{sec:results} and conclude with a discussion in Sec.~\ref{sec:discussion}. 
	
	\section{The Fibonacci code} \label{sec:code}
	
\begin{figure}
		\centering
		\begin{subfigure}{.58\linewidth}
			\centering
			\includegraphics[width=1\linewidth]{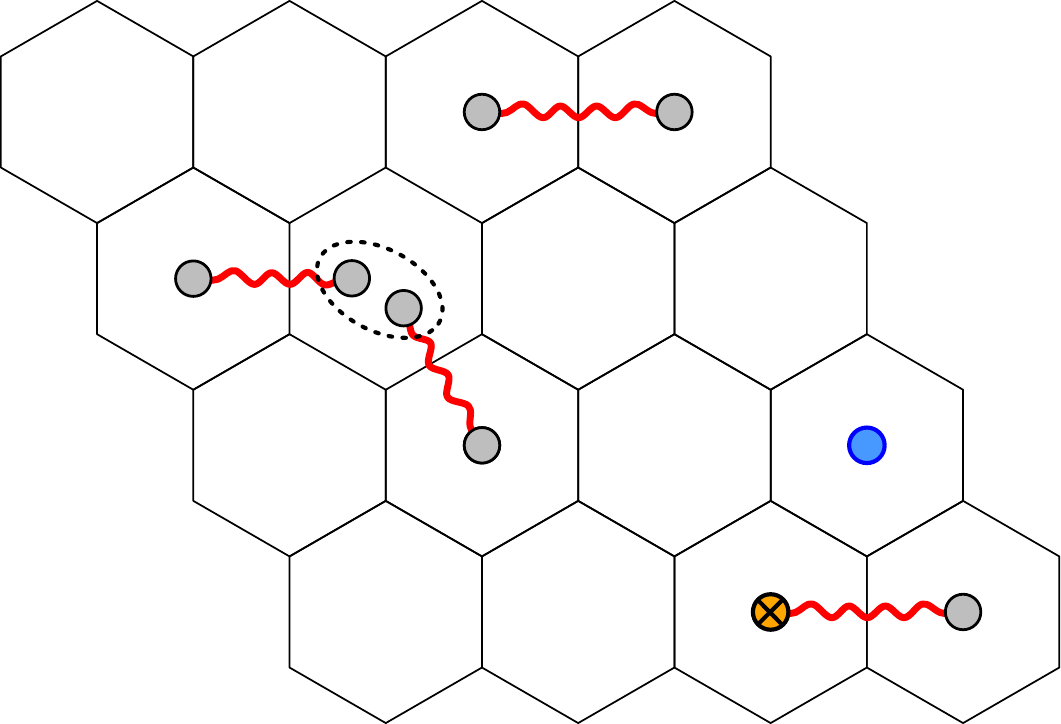}
			\caption{}
		\end{subfigure}
		\hfill
		\begin{subfigure}{.4\linewidth}
			\centering
			\includegraphics[width=.95\linewidth]{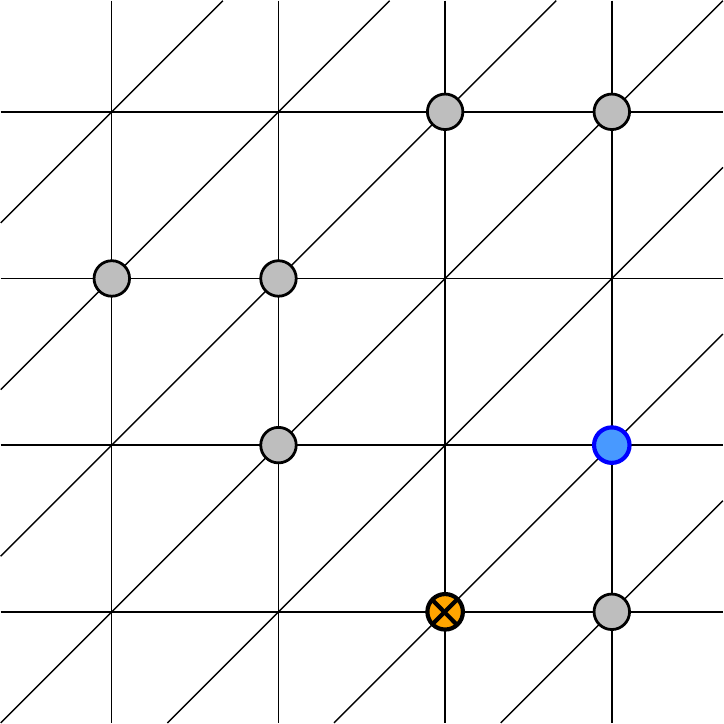}
			\caption{}
		\end{subfigure}
		\caption{(a) Pair-creation events creating anyonic excitations in neighboring tiles. The dotted ellipse represents a collapse to the total charge of the anyons it contains. A ghost defect is shown in blue, a missing defect is highlighted in orange with a cross. (b) The outcome of this noise process represented on the decoding graph. Note that the missing defect (highlighted in orange) will (by definition) not be visible in the syndrome.}
		\label{fig:pair-creation}
	\end{figure}

	We consider a two-dimensional model comprised of hexagonal tiles laid out on the surface of a torus. The resulting geometry can be represented as an $ L \times L $ hexagonal lattice with periodic boundary conditions in both directions (Fig.~\ref{fig:pair-creation}). Each of these hexagonal tiles can contain an excitation known as a \emph{Fibonacci anyon}.
	
	Anyons are point-like quasi-particle excitations which can be characterized algebraically in terms of a unitary modular tensor category (UMTC). A thorough description of anyon models using UMTCs goes beyond the scope of this work, however, some details are given in Sec.~\ref{sec:anyons}. 
	For now, it is sufficient to state that an anyon model specifies a set of anyon labels, also referred to as particle types, which can fuse according to a specific set of fusion rules.
	The Fibonacci anyon model considered in this work contains two labels, $ \1 $ and $ \tau $, which obey the fusion rules
	\begin{equation}\label{eq:fusion_rules}
		\1 \times \1 = \1\,, \quad \1 \times \tau = \tau \times \1 = \tau\,, \quad \tau \times \tau = \1 + \tau\,.
	\end{equation}
	
\begin{figure}
		\centering
		\includegraphics[width=0.7\linewidth]{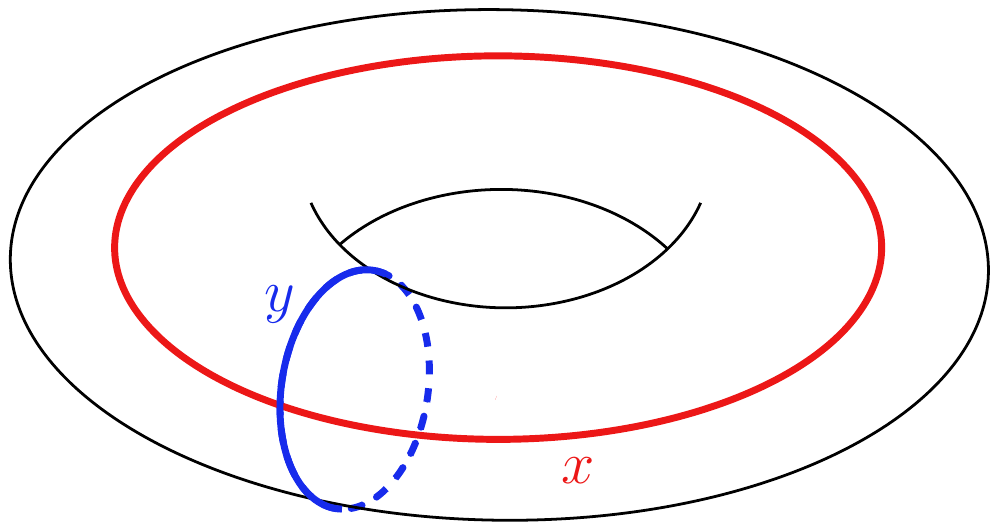}
		\caption{The two homologically non-trivial cycles on a torus.}
		\label{fig:torus}
	\end{figure}

	In general, one can associate a vector space to a given set of anyons, where the basis vectors are labeled by the different ways in which the anyons can fuse. This fusion space has a topological degeneracy, and can therefore be used to robustly encode quantum information. In particular, for the Fibonacci anyon model the anyonic vacuum on a two-dimensional torus has a twofold degeneracy \cite{pfeifer2012translation}. Starting from our two-dimensional model, we can therefore define an error correcting code whose code space is identified with the anyonic vacuum on the torus and which encodes a single logical qubit.
	A basis for this code space can be defined using Wilson line operators along the homologically non-trivial cycles $ x $ and $ y $ shown in \figref{fig:torus}:
	\begin{equation}\label{eq:basis}
	    \begin{array}{l}
	        W^a_x \ket{\1}_x = \frac{S_{a\1}}{S_{\1 \1}} \ket{\1}_x\,,\\
		 W^a_x \ket{\tau}_x = \frac{S_{a\tau}}{S_{\1 \tau}} \ket{\tau}_x\,, 
	    \end{array}
		\qquad  a \in \{\1,\tau\} \,,
	\end{equation}
	We note that a different basis, $ \{\ket{0}_y, \ket{1}_y\} $, can be defined analogously by swapping the $ x $ and $ y $ labels above, where the two bases are related through the modular $ S $ matrix $ S_{ab} = \!\!\phantom{a}_y\!\braket{a|b}_x$.
	For the Fibonacci anyon model its numerical values are
	\begin{equation}\label{eq:S}
		S = \dfrac{1}{\sqrt{1+\phi^2}} 
		\begin{pmatrix}
			1 & \phi \\
			\phi & -1
		\end{pmatrix}\,,
	\end{equation}
	where $ \phi = \dfrac{1 + \sqrt{5}}{2} $.

	
	The action of the mapping class group on the anyonic vacuum then corresponds to unitary operations on the code space. For the Fibonacci category, any logical unitary operator can be realized in this way, up to arbitrary precision \cite{freedman2002modular}. Therefore, the quantum error correcting code defined above natively supports universal quantum computation. 
	
	Errors in this code appear as spurious anyonic excitations, which can corrupt the encoded information if their world lines between creation and re-annihilation are topologically non-trivial, i.e., form a non-contractible cycle \footnote{Note that a pair of Fibonacci anyons can also fuse to a single non-trivial anyon when one member of the pair has been transported along a non-trivial cycle. Since the resulting state is no longer in the code space, this does not constitute a logical operation on the encoded information. However, one can show that the encoded information is irrevocably lost in case of such an event \cite{wootton2014error}.}. The objective of error correction is then to systematically remove these spurious excitations, without corrupting the quantum memory in the process.
	This correction is performed in an active and continuous manner, and can be broken down into a series of discrete steps. At each step, a suitable recovery operation is performed based on a measured list of positions and types of the excitations, called the error syndrome.
	
	We conclude this section by noting that the numerical simulation of the error-correction process requires the introduction of some additional manipulations on fusion states of multiple Fibonacci anyons. As these are technical details that do not contribute to the intuition of the procedure, we defer their definition to Sec.~\ref{sec:anyons}.

	\section{Noise model and correctability} \label{sec:noise}
	
	Having defined our model and code space, we now turn to the description of the noise model used in our simulations.
	We model continuous active error correction in our Fibonacci code as a sequence of time steps, where each time step itself consists of three parts: the application of pair-creation noise, faulty syndrome measurement, and error correction respectively. 
	
	At each time step, first, for each edge of the hexagonal lattice graph a pair of anyons is created across this edge with a probability $ p $. Immediately after each pair creation event, the resulting charge in the two affected tiles is sampled, effectively collapsing all superpositions of anyonic charge within each tile to either $\1$ or $\tau$. After the pair creation noise has been applied, faulty syndrome extraction is simulated by generating a list of the anyon charge in all tiles, and flipping each outcome individually with a probability $ q $. In addition to the charges that are correctly detected, the resulting faulty syndrome can contain both \enquote{ghost defects} (indicating a non-trivial charge when none is truly present) and \enquote{missing defects} (failing to report a true non-trivial charge).
	Finally, this faulty syndrome is passed to a decoder, introduced in the following section, which then performs a set of local operations based on the current (and past) syndrome information in an attempt to move the system back towards the initial state.
	
	After each time step, the current state of the system is copied and it is checked whether it is still correctable. This is done by passing the copy to a clustering decoder \cite{brell2014thermalization, burton2017classical, schotte2022quantum} and simulating a decoding procedure with perfect syndrome measurements starting from this given initial state. If this perfect decoding is successful, the memory is considered intact and the simulation is continued. If perfect decoding is unsuccessful, the memory is considered corrupted and the simulation is aborted. The memory lifetime is then defined as the number of time steps after which a perfect clustering decoder can no longer successfully restore the initial state.
	
	For a given pair of tiles which share an edge, the process of pair creation across this edge corresponds to the matrix elements
	\begin{equation}
		\bra{a'\!,b'\!;c'}U_{\text{pc}}\ket{a,b;c} = \delta_{c,c'} F^{a a 1}_{\tau\tau a'} F^{a'\!\tau a}_{b \, c \, b'}\,,
	\end{equation}
	where we have used the $ F $-symbols of the Fibonacci category, given in \eqref{eq:Fib_F_symbols_nontriv}. Here, $ \ket{a,b;c} $ represents the state where the affected tiles have anyon charges $ a $ and $ b $, respectively, with total charge $ c $.  This then defines the probability distribution according to which outcomes $ a' $ and $ b' $ are sampled. 
	Since our noise model does not allow any superposition in the anyonic charge of individual tiles, it should be considered semi-classical rather than fully quantum-mechanical. Note, however, that this does not render our model completely classical. Indeed, superpositions in the total charge $ c $ of the affected tiles are an inherent part of the state evolution that cannot be captured faithfully by any classical probabilistic process.  
	Furthermore, while the extreme decoherence assumption for the anyon charge in individual tiles greatly simplifies the numerical simulation outlined in this work, it was argued in Ref.~\cite{brell2014thermalization} that this decoherence is unlikely to have any tangible influence on the observed memory lifetimes, as the essential topological nature of the noise processes is still captured correctly.
	
	We note that this type of noise can be understood as originating from the connection to a thermal bath with inverse temperature  $ \beta = 1/(k_B T) $ determined by the error rate $ p $ through the relation
	\begin{equation}\label{key}
		\dfrac{p}{1-p} = \e^{-\beta \Delta}\,.
	\end{equation}
	Here, $ \Delta $ represents the energy required to create a pair of anyonic excitations and place them in neighboring tiles.
	
	To conclude this section, we emphasize that in the case of non-Abelian error correction, even decoding with perfect syndrome measurements is still an inherently stochastic procedure due to the indeterminacy of anyonic charge measurements. This means that perfect decoding can sometimes either be successful or unsuccessful even starting from the same initial state. Our definition of the memory lifetime therefore simply corresponds to a statistical estimate of the actual memory lifetime. 
	Furthermore, it is not known which decoder is optimal for the Fibonacci code. Hence, the choice for the clustering decoder to verify the correctability of states is, in a way, an arbitrary one. This choice, however, is motivated by the recent discovery that the clustering decoder yields high thresholds for a related error correcting code exhibiting doubled Fibonacci anyonic excitations, and performed significantly better in this context than decoders based on a perfect matching strategy \cite{schotte2022quantum}.
	In any case, one should keep in mind that the memory lifetime as defined above, does not represent the true memory lifetime. Instead the sub-optimal verification process entails that it merely provides us with a lower bound on the true value.

	\section{Harrington's cellular automaton decoder} \label{sec:decoder}
	The model described above is paired with a decoder which is a straight-forward adaptation of the cellular automaton decoder introduced in Ref.~\cite{harrington}.
	Previously, this decoder has also been used for a similar phenomenological model of Ising anyons in Ref.~\cite{dauphinais2017fault}, where the existence of an error correction threshold was proven analytically.
	At each time step during the error correction simulation, based on the reported measurement outcomes in the faulty syndrome, the decoding algorithm will apply local transition rules to fuse neighboring anyons or to move anyons to neighboring tiles.
	
\begin{figure}[h]
		\centering
		\begin{subfigure}{.45\linewidth}
			\includegraphics[width=\linewidth]{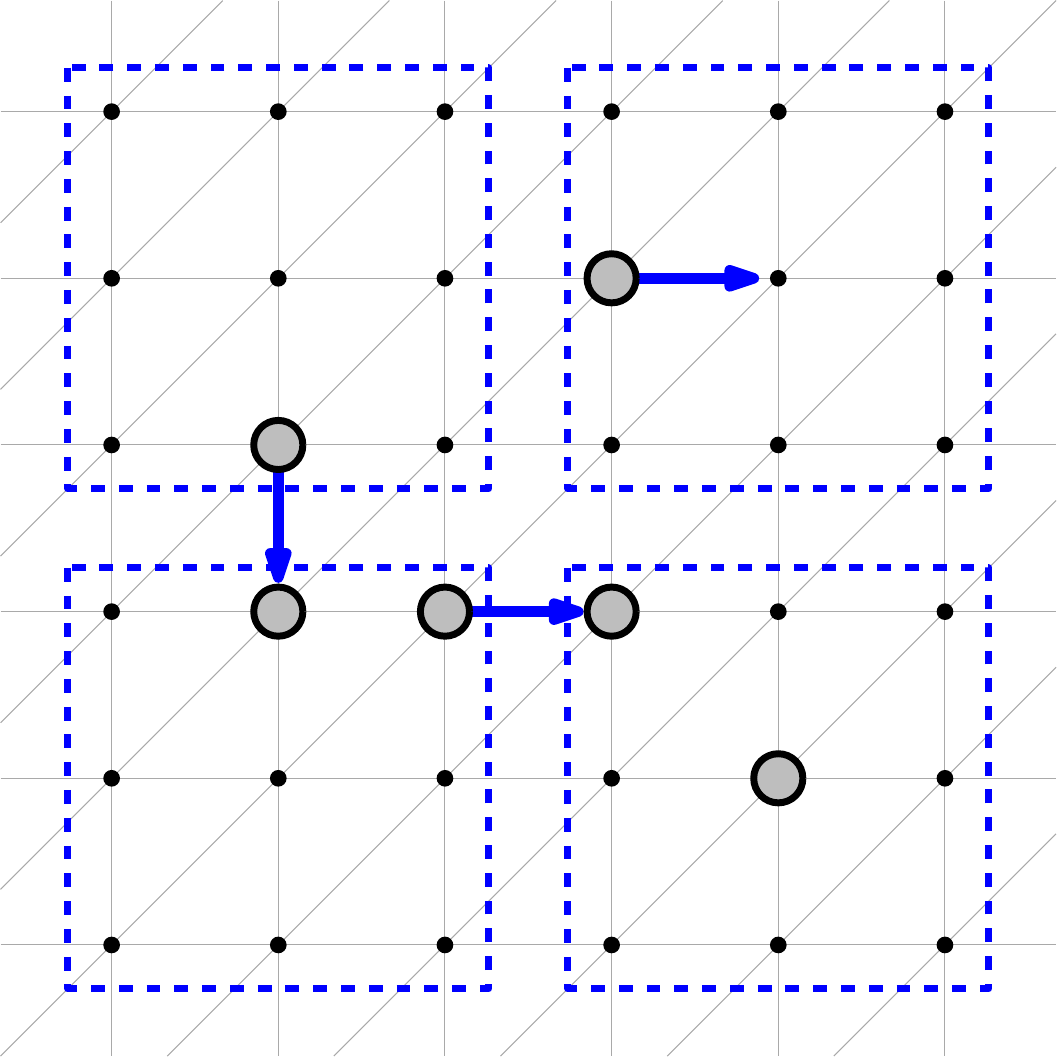}
			\caption{}
			\label{fig:transition_1}
		\end{subfigure}
		\hfill
		\begin{subfigure}{.45\linewidth}
			\includegraphics[width=\linewidth]{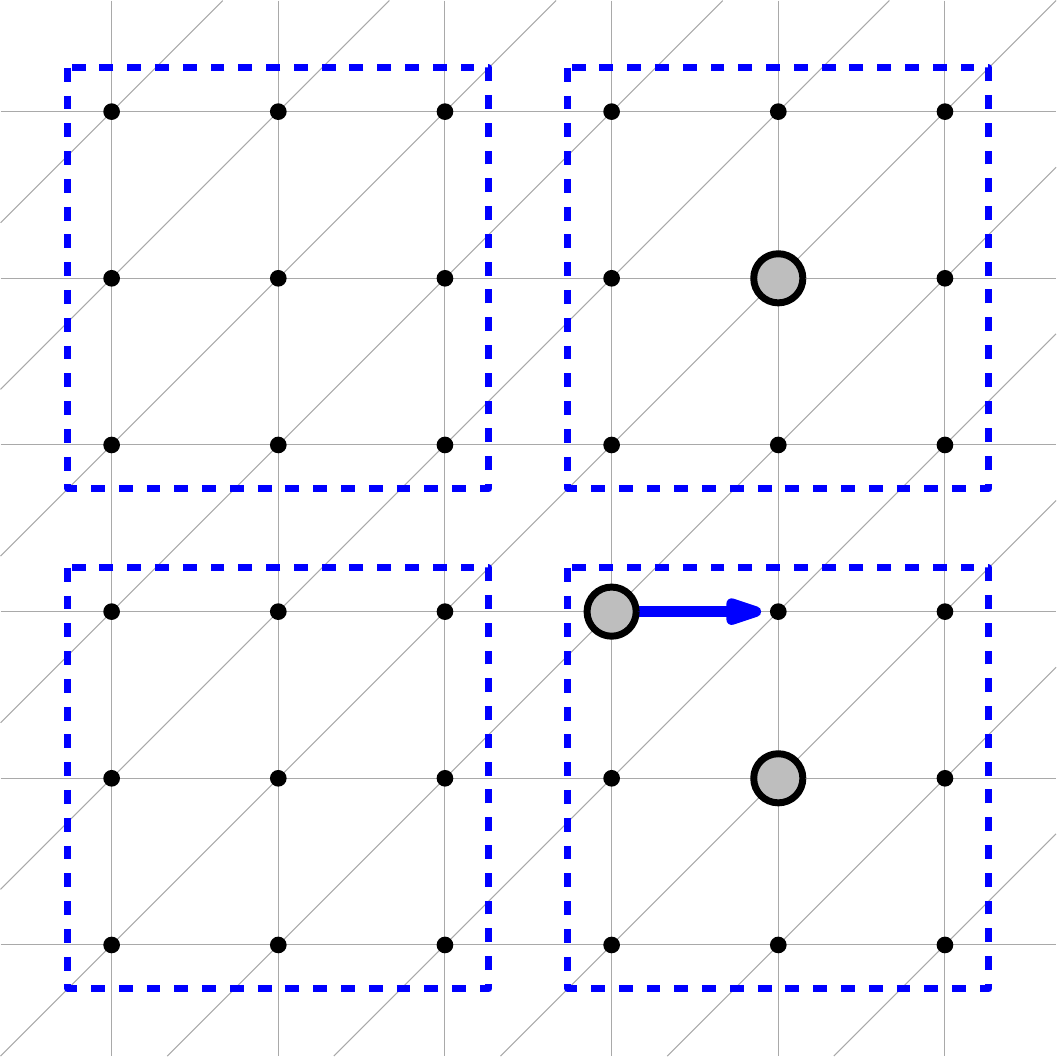}
			\caption{}
			\label{fig:transition_2}
		\end{subfigure}
		\hfill
		\begin{subfigure}{.45\linewidth}
			\includegraphics[width=\linewidth]{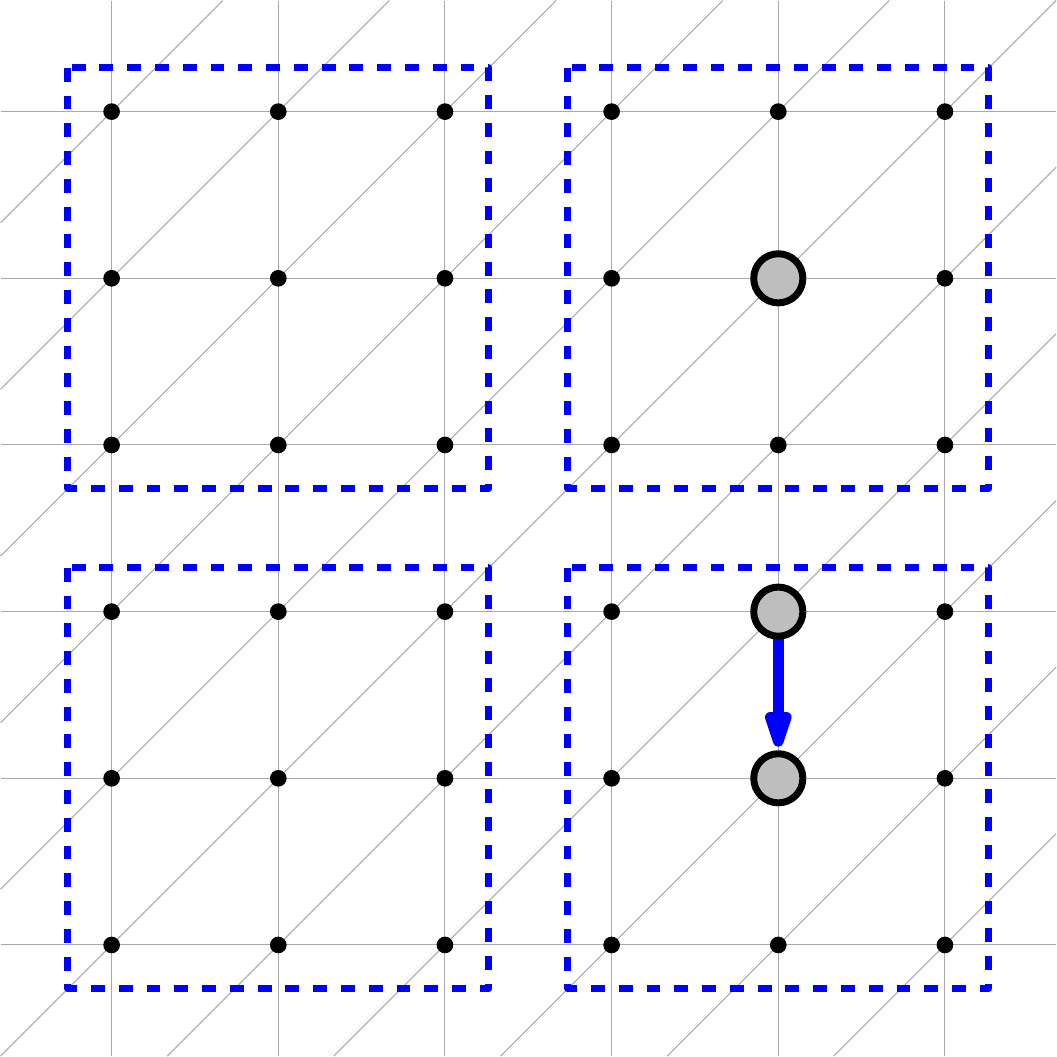}
			\caption{}
			\label{fig:transition_3}
		\end{subfigure}
		\hfill
		\begin{subfigure}{.45\linewidth}
			\includegraphics[width=\linewidth]{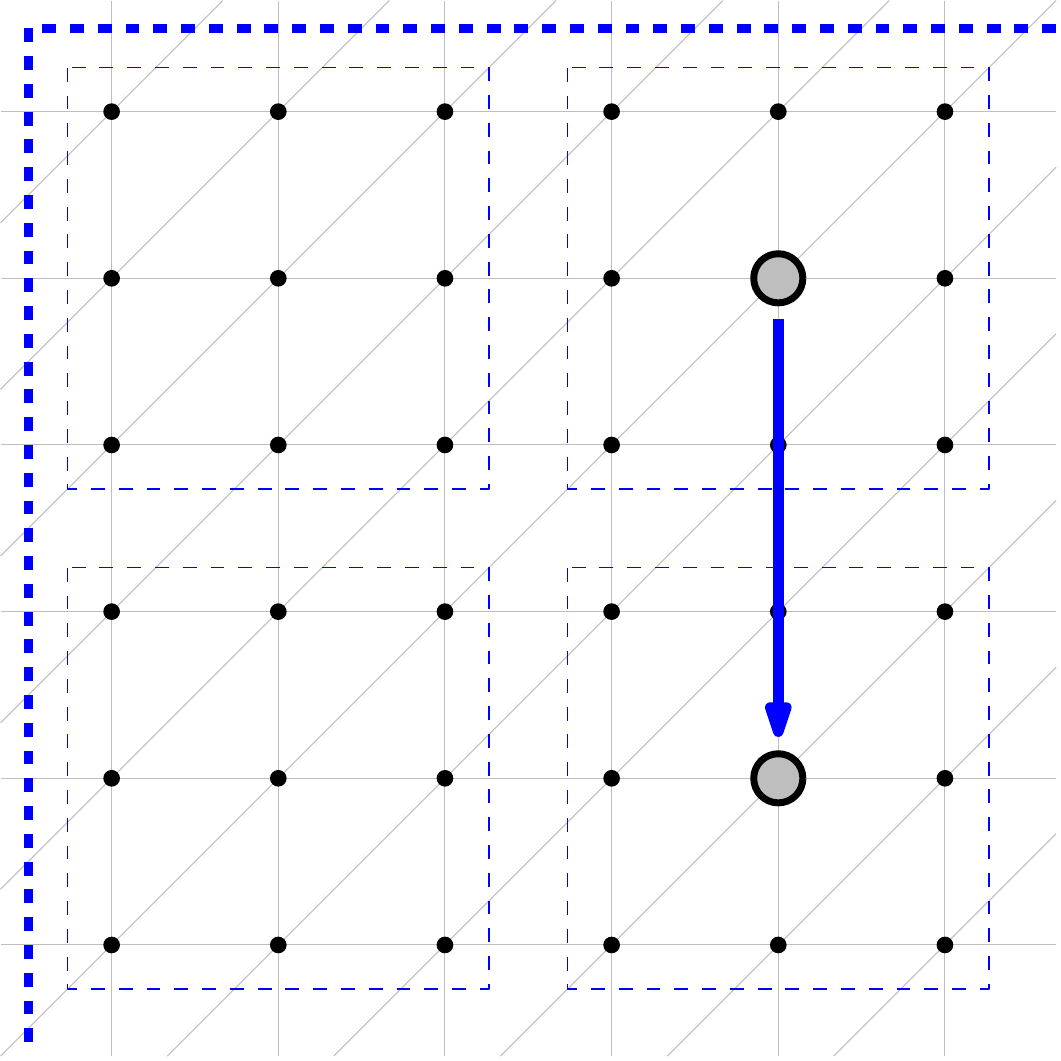}
			\caption{}
			\label{fig:transition_4}
		\end{subfigure}
		\caption{Illustration of the transition rules on the decoding graph. The gray disks represent non-trivial syndromes, and the blue arrows represent the actions suggested by the decoder. The blue dotted lines represent the $3\times3$ colonies. (a-c) show a sequence of level-0 transition rules and possible outcomes of those actions. In (d) non-trivial anyons have been transported to two neighboring colony centers, the blue arrow represent a level-1 transition which could be applied at the end of the work period.}
		\label{fig:transition_rules}
	\end{figure}

	Intuitively these transition rules work as follows. The lattice is divided into square colonies of size $Q \times Q$. At each time step, the transition rules will attempt to fuse neighboring non-trivial anyons, as observed in the faulty syndrome. If a non-trivial anyon has no neighbors, the transition rules will move it to the center of its colony. 
	At larger timescales, higher-level transition rules are applied on a renormalized lattice where anyons located at colony centers will be fused with anyons at neighboring colony centers, or moved toward the center of their respective super-colonies, which consist of $Q \times Q$ colonies. This renormalization scheme is then continued at higher levels until eventually the $ Q^n \times Q^n $ super-colony covers the entire lattice for some integer $ n $. 
	To ensure the latter is possible, we will always assume that the linear lattice size satisfies $L= Q^n$ for some integer $ n $.
	An example of these processes is shown in \figref{fig:transition_rules}.
	
	To describe the action of the decoding algorithm more precisely, we will define its action at different renormalization levels $ k $. 
	The level-0 transition rules are those already discussed above and are applied at every time step based on the reported faulty syndrome obtained from the most recent round of faulty measurements. The transition rules are applied to one location at a time and take into consideration only the anyon content of that site and of its eight neighbors. 
	A detailed definition of these rules is given in App. \ref{sec:transition_rules}.
	When an anyon is moved from a site $ l $ to a neighboring site $ l' $, the (true) anyon content of site $ l $ is fused with that of site $ l' $ and the resulting charge is placed on site $ l' $ while the charge of site $ l $ is restored to the vacuum. This happens irrespective of whether or not the syndrome for both sites was correct. 
	Hence, when the decoder attempts to move a ghost defect (a trivial charge misidentified as a non-trivial one) to a neighboring site, this process does not create additional excitations. This does not, however, mean that mistaking a trivial charge for a non-trivial one has no negative consequences. Indeed, these wrong syndromes may cause the decoder to stretch out existing errors.

	The level-1 transition rules are not applied in every time step, but only when $ t $ is a multiple of a parameter called the \emph{work period}, which we will denote by $ U $. We require that $ U = b^2 $ for some positive integer $ b $. 
	One should think of $ U $ as the time scale at which a coarse-graining is performed. 
	Level-1 transition rules act at on a coarse-grained lattice where the sites correspond to the centers of the level-0 colonies, and these are grouped into level-1 colonies of size $ Q^2 \times Q^2 $. 
	Hence, the actions determined by the level-1 transition rules involve pairs of level-0 colony centers separated by a distance $ Q $. An example of such a move is provided in \figref{fig:transition_4}.	
	The transition rules themselves are nearly identical to the level-0 rules, but are based on two sets of level-1 syndromes $ s_{1,c} $ and $ s_{1,n} $ (defined below) rather than one. For a site $ l $ (which is a level-0 colony center), the transition rules use $ s_{1,c}(l) $ as the anyon content of site, while the anyon content of its neighbors (that is, the neighboring level-0 colony centers) is taken to be $ s_{1,n}(l') $.	
	
\begin{figure}[h]
		\vspace{.5cm}
		\centering
		\includegraphics[width=\linewidth]{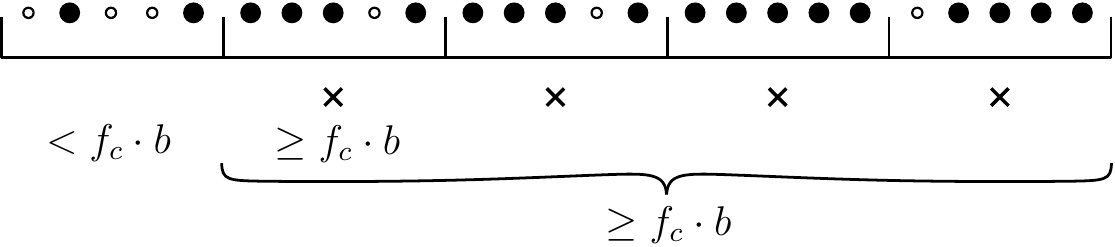} 
		\caption{A visual example of the coarse-grained counting procedure to determine the level-1 syndrome for a level-0 colony center. The row of dots represent $ U $ time steps, divided in $ b $ intervals of size $ b $. The time steps during which a non-trivial measurement outcome was reported are indicated by the colored dots. The crosses in the second row indicate in which intervals the fraction of non-trivial measurement outcomes is equal to or higher than $ f_c $.  }
		\label{fig:binning}
	\end{figure}

	The definitions of the level-1 syndromes $ s_{1,c} $ and $ s_{1,n} $ require a pair of variables $ f_c, f_n \in [0,1]$. 
	Intuitively these variables serve as detection thresholds for the level-1 syndromes by determining the fraction of measurements that must return a non-trivial outcome at a site before it qualifies as a non-trivial level-1 syndrome. 
	The proper definition, however, is slightly more complicated and uses a coarse-grained counting method. 
	Below, we give the precise definition of $ s_{1,c} $,  the definition of $ s_{1,n} $ is entirely analogous (using $ f_n $ instead of $ f_c $). 
	We start by dividing the work period $ U = b^2 $, into $ b $ intervals of $ b $ time steps each.
	For each of these intervals, we say a non-trivial syndrome is present at a colony center $ l $ if a non-trivial charge was reported there for at least $ f_c b $ of the $ b $ time steps in the interval. 
	When at least $ f_c b $ of the $ b $ intervals have a non-trivial syndrome, $ s_{1,c}(l) $ is set to one. 
	A visual example of this coarse-grained counting procedure is shown in \figref{fig:binning}.
	
	The motivation for using two types of syndromes for $ k>0 $ is as follows. Suppose that an error spans across two neighboring colonies, which we will label $ \rho $ and $ \rho' $. The level-0 transition rules will transport all resulting anyons to the respective colony centers, where they can now be acted upon by level-1 transition rules at the end of the work period. Imagine that a non-trivial anyon is now present at both colony centers. When considering the level-1 transition rules acting on $ \rho $, there are four possible scenarios for the syndromes $ s_{1,c}(\rho) $ and $ s_{1,n}(\rho') $. In case $ s_{1,c}(\rho) = 0 $, the transition rules act trivially on $ \rho $. If both $ s_{1,c}(\rho) = 1 $ and $ s_{1,n}(\rho') = 1 $ then the transition rules will be applied correctly and the anyons will be fused. 
	However, if $ s_{1,c}(\rho) = 1 $ but $ s_{1,n}(\rho') = 0 $, the transition rules may move the anyon in $ \rho $ away from $ \rho' $, thereby increasing the weight of the error.
	Hence, it is desirable to set $ f_c > f_n $ to decrease the odds that when a level-k syndrome reports an non-trivial anyon at a colony center, the level-k syndrome for its neighbors are false negatives. 
	We must be careful not to set $ f_c $ too high or $ f_n $ too low, however. If we choose $ f_c $ to high, low-weight errors could cause $ s_{1,c} $ to never report any non-trivial charges, delaying any necessary corrections. Similarly, setting $ f_n $ too low will result in low-weight error triggering many false positives for $ s_{1,n} $, which can cause the decoder to make wrong decisions. 
	
	Level-$ k $ transition rules are applied when $ t \mod U^k = 0 $. They operate on a renormalized lattice that uses the centers of level-$ (k-1) $ colonies as sites, and groups these into level-$ k $ colonies of size $ Q^k \times Q^k $. The level-$ k $ syndromes $ s_{k,c} $ and $ s_{k,n} $ are determined by the coarse-grained counting method described above, using $ b^k $ intervals of $ b^k $ time steps each. 
	For linear system size $ L $, k ranges from 0 to $ k_{\text{max}} = \log_Q(L) $.

	It is important to note that non-Abelian anyons do not allow for instantaneous moves. 
	Indeed, while one can construct a unitary string operator for Abelian anyons, no such operator can be constructed for the non-Abelian case. 
	This discrepancy can be traced back to the fact that fusion outcomes are non-deterministic for non-Abelian anyons, implying it is not possible to move an non-Abelian anyon by annihilating it with one member of a particle-antiparticle pair (as is done in e.g., the surface code).
	
	Therefore, the actions determined by level-$k$ transition rules, for $k>0$ cannot be applied withing a single time step. Instead, they will be broken up into a sequence of moves involving only pairs of neighboring sites which will be applied in $Q^k$ consecutive time steps. 
	We further limit the model by requiring that the number recovery operations affecting a single tile in the lattice (or site in the decoding graph), is no greater than one in each time step. 
	This allows all recovery operations applied in one time step to be performed in parallel. 
	Hence, we must define a hierarchy determining which actions (moves or fusions between neighboring tiles) get prioritized based on the renormalization level from which they originated. In our case, it was opted to always prioritize correction processes from the highest renormalization level \footnote{Note that if one were to prioritize the level-0 corrections, higher-level correction could never be completed, as they would be undone immediately after their first action is applied.}
	
	It was argued in \cite{dauphinais2017fault} that the prohibition of instantaneous corrections would likely not influence the threshold behavior other than slightly lowering the memory lifetimes relative to a hypothetical case where this restriction is dropped. We explicitly verify this claim for our Fibonacci model below in Sec.~\ref{sec:results}.
	
	\begin{figure}[h]
		\centering
		\hfill
		\begin{subfigure}{.45\linewidth}
			\includegraphics[width=\linewidth]{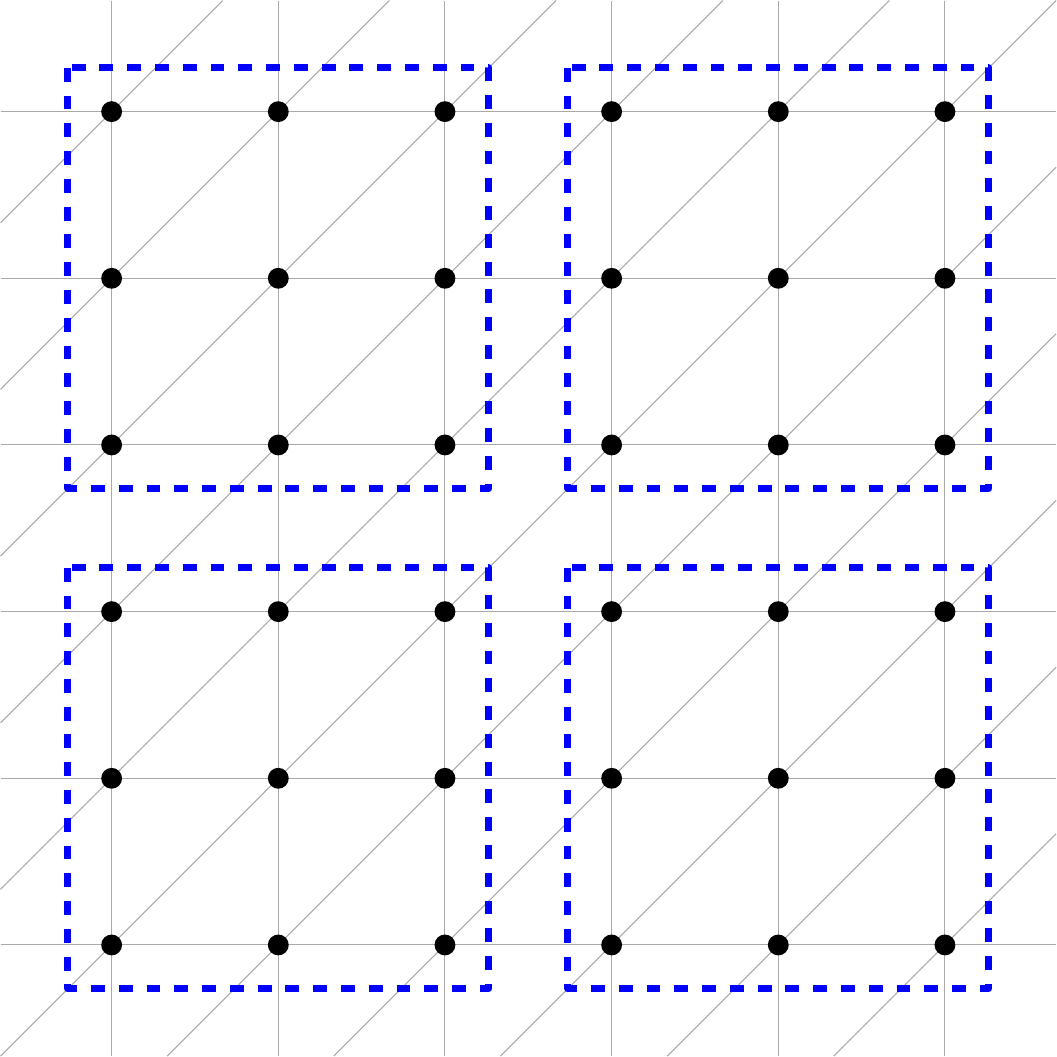}
			\caption{}
			\label{fig:colonies1}
		\end{subfigure}
		\hfill
		\begin{subfigure}{.45\linewidth}
			\includegraphics[width=\linewidth]{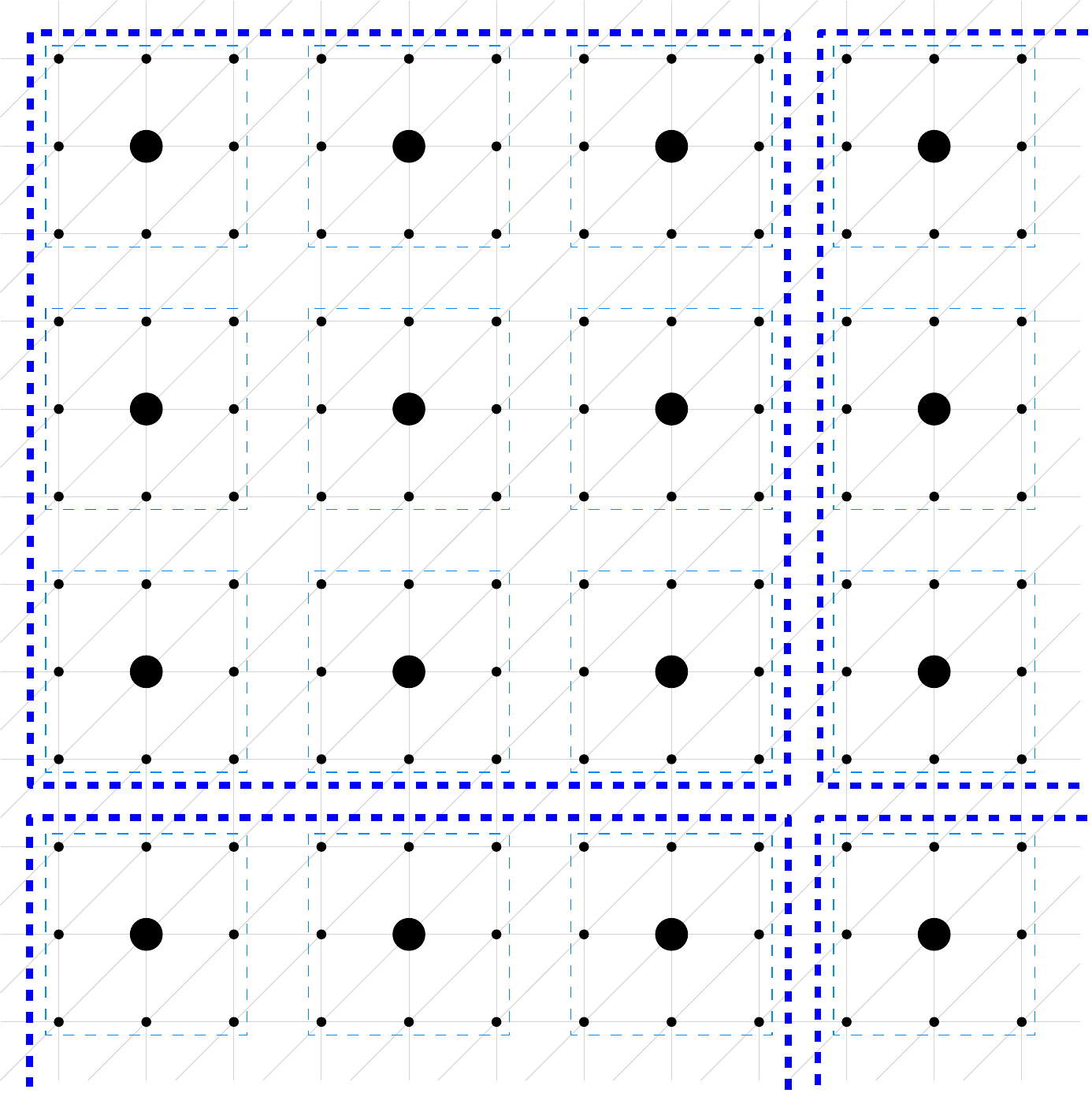}
			\caption{}
			\label{fig:colonies2}
		\end{subfigure}
		\hfill
		\begin{subfigure}{.45\linewidth}
			\includegraphics[width=\linewidth]{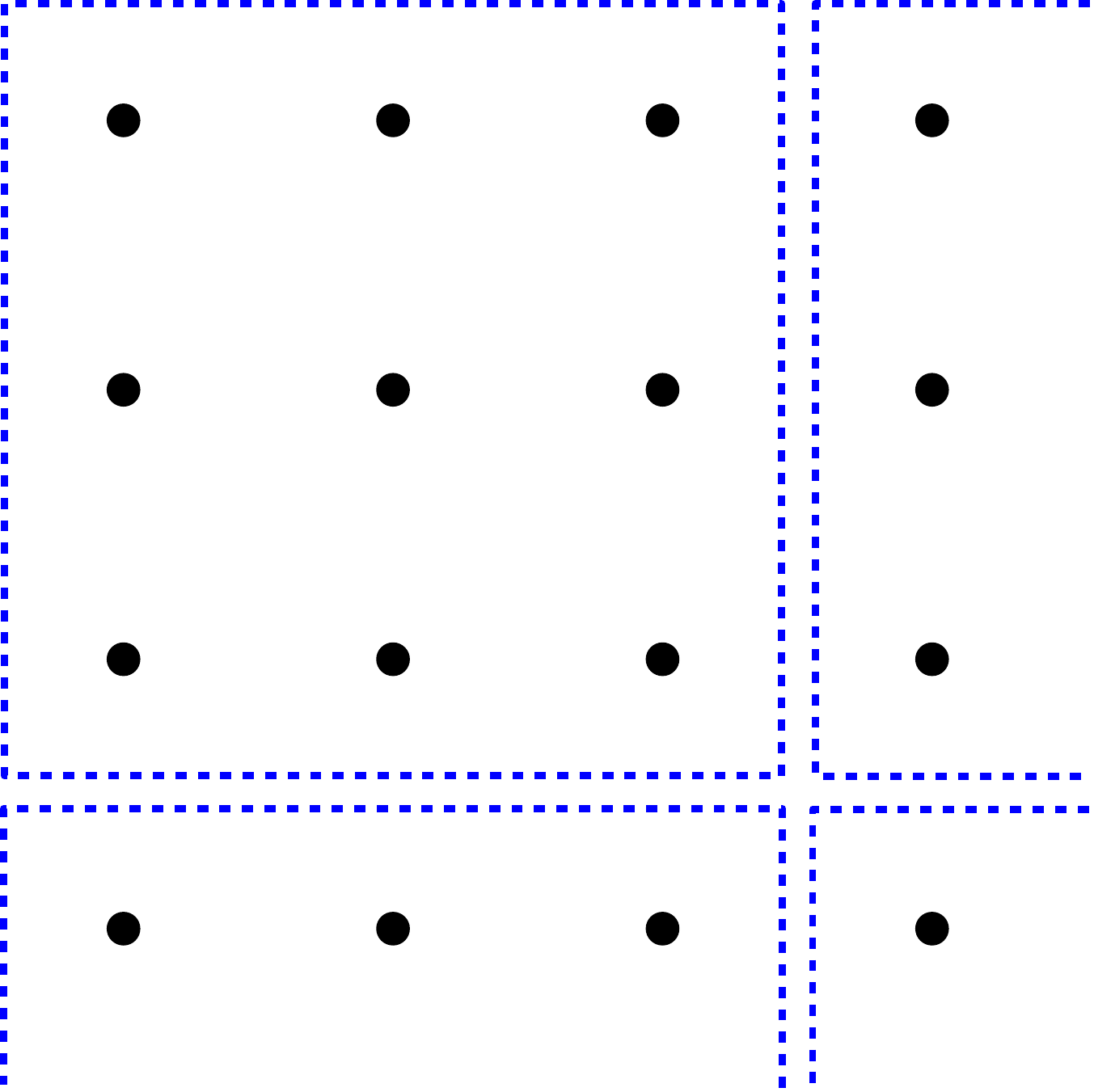}
			\caption{}
			\label{fig:colonies3}
		\end{subfigure}
		\hfill
		\caption{(a) Level-0 colonies of size $Q\times Q$. (b) Level-1 colonies defined as $Q\times Q$ level-0 colonies. (c) Renormalized lattice used for the level-1 transition rules. }
		\label{fig:colonies}
	\end{figure}

	\section{Outline of the simulation} \label{sec:simulation}
	
	The goal of this work is to numerically determine a fault-tolerant error threshold for the error correcting code defined in Sec.~\ref{sec:code} with pair-creation noise and measurement noise as outlined in Sec.~\ref{sec:noise}, and with the cellular automaton decoder introduced in Sec.~\ref{sec:decoder}.
	This is achieved by performing Monte-Carlo simulations to determine the average memory lifetime for a range of system sizes and error rates. 
	These results then allow one to estimate the value of the error threshold. 
	
	A single Monte-Carlo sample (with some fixed values for the noise strength $p$ and the measurement error rate $q$) is obtained as follows.
	First, the state of the system is initialized as a ground state (i.e.: containing no anyons).
	Then a sequence of time steps is performed consisting of the application of pair-creation noise with rate $p$, a round of faulty syndrome measurements with error probability $q$, and finally a sequence of recovery operations. 
	At the end of each time step, it is verified whether or not the state is considered correctable, according to the criteria specified in Sec.~\ref{sec:noise}. 
	This sequence of time steps is continued until one of the following three outcomes occurs: (1) The largest connected group of anyons grows too large, rendering its classical simulation intractable \footnote{Note that such cases are likely to correspond configurations in which the initial state cannot be recovered.}; 
	(2) A noise process or recovery operation induces a logical error by fusing a pair of anyons along a path that forms a non-contractible loop when combined with their fusion tree;
	(3) The verification procedure at the end of a time step fails. 
	The memory lifetime is then set as the number of time steps that were completed.
	The course of a single Monte-Carlo sample in the simulation is summarized as pseudo-code in Alg. \ref{alg:decoder}.
	
	\begin{algorithm}[H]
		\caption{Numerical simulation} \label{alg:decoder}
		\begin{algorithmic}
			\State initialize state
			\State $t = 0$
			\While{correctable with clustering decoder $\And$ no logical errors made}  
			\State $t \gets t+1$
			
			\State apply pair-creation noise
			\State perform faulty measurements
			\For{$k = 0: k_{\text{max}} $}
			\If{$t\!\mod U^k = 0$}
			\State update level-$ k $ syndromes
			\State apply level-$ k $ transition rules
			\EndIf
			\EndFor										
			\EndWhile
			\State memory lifetime = t
		\end{algorithmic}
	\end{algorithm}
	
\section{Numerical results}\label{sec:results}
	
\begin{figure*}[t]
	    \centering
    	\begin{subfigure}[t]{.49\textwidth}
    		\centering
    		\includegraphics[trim={.8cm 7.3cm .5cm 6cm},clip,width=1.0\linewidth]{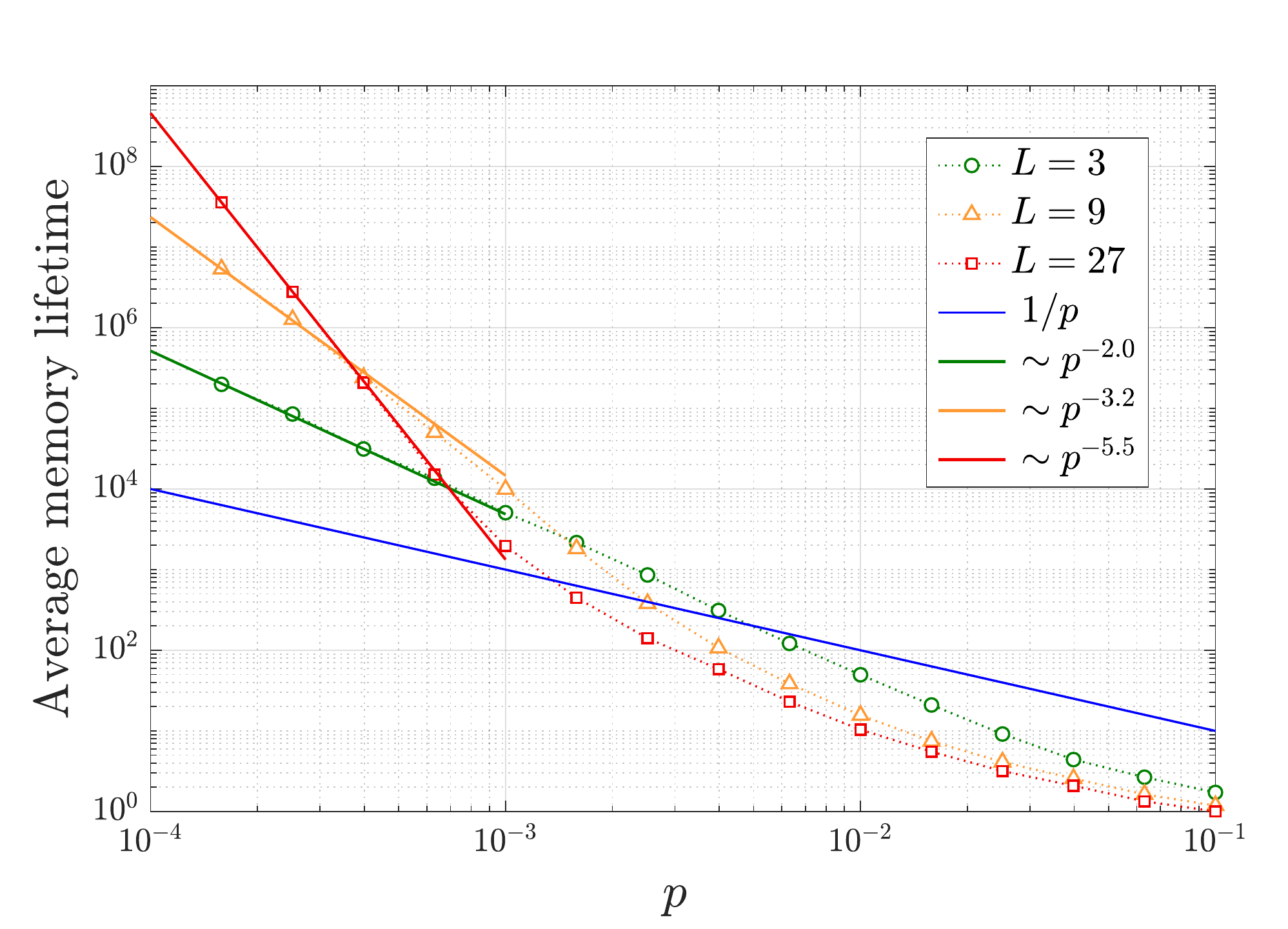}
    		\caption{Average memory lifetime in function of the error strength, with $ p = q $, for various system sizes. Each data point represents the average over 1000 Monte Carlo samples. The blue line shows the coherence time of a single physical qubit. 
    		The average memory lifetimes for $ p \leq 10^{-3}  $ were fitted to a function of the form $ f(p) \sim p^{-a} $. The results for $ L = 3 $, $ L = 9 $ and $ L = 27 $ are shown as the green, yellow and red lines respectively. }
    		\label{fig:results}
    	\end{subfigure}
	    \begin{subfigure}[t]{.49\textwidth}
    		\centering
    		\includegraphics[trim={.8cm 7.3cm .5cm 6cm},clip,width=1.0\linewidth]{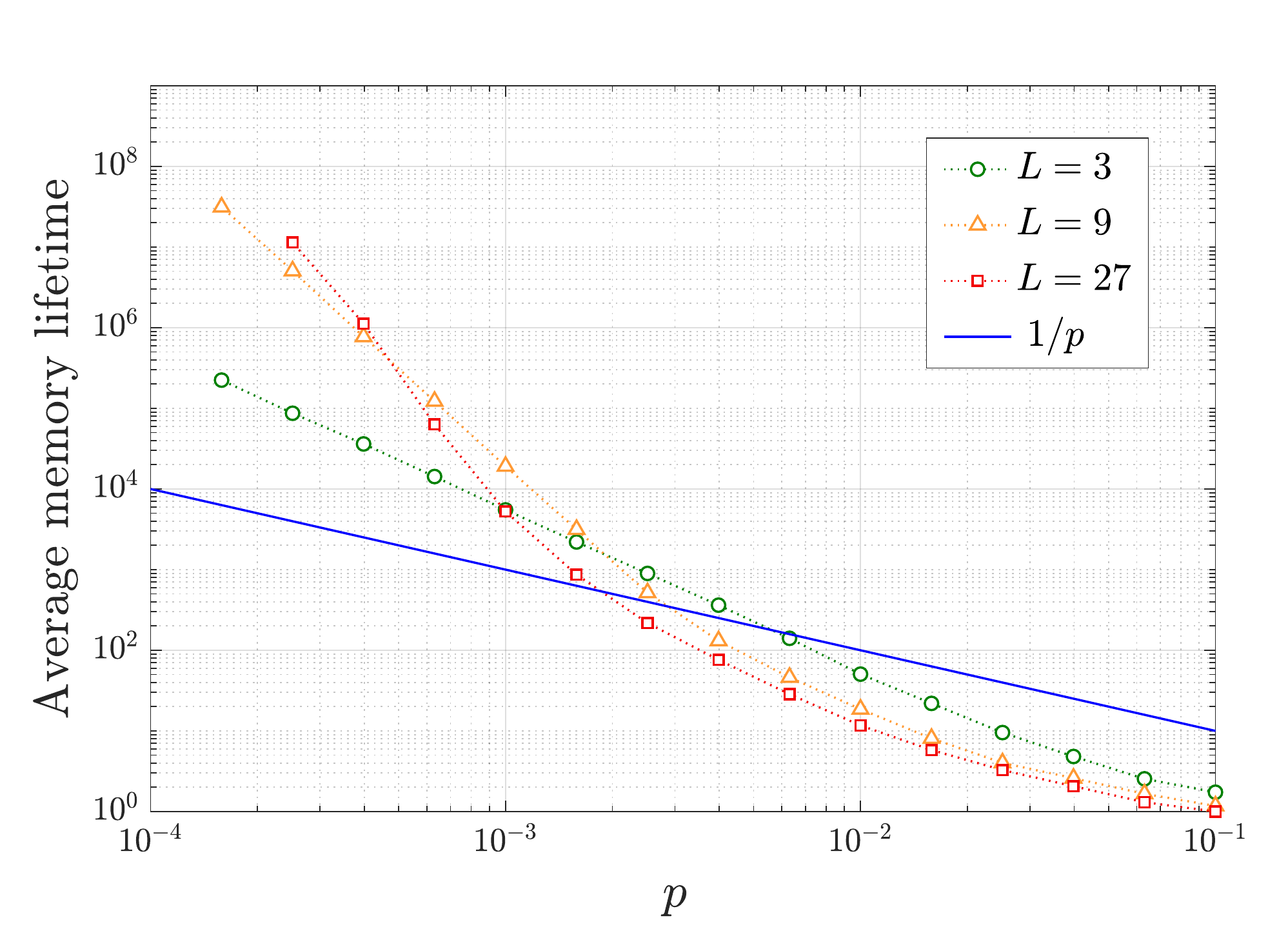}
    		\caption{Average lifetime in function of the error strength with $ p = q $ for various system sizes and (unphysical) \emph{instantaneous} corrections. Each data point represents the average over 1000 Monte Carlo samples. The blue line shows the coherence time of a single physical qubit. }
    		\label{fig:results_instant}
    	\end{subfigure}
	    \caption{}
	    \label{fig:results_both}
	\end{figure*}

	The Monte Carlo simulation described above were performed for various system sizes with $ p = q  $. 
	The following parameters were used:
	\begin{align*}
		Q &= 3\,,\\
		b &= 7\,,\\
		F_c &= 0.7\,,\\
		F_n &= 0.2\,. \\ 
	\end{align*}
	The resulting average memory lifetimes for $ L = 3 $, $ L = 9 $ and $ L = 27 $ are shown below in \figref{fig:results}.
	
	These results clearly indicate that the code presented in this work is indeed fault-tolerant. Furthermore, while the current data is not sufficient to demonstrate a clear-cut fault-tolerant threshold, it still exhibits threshold behavior and is remarkably similar to the results previously obtained for the toric code \cite{harrington} and the Ising topological code \cite{dauphinais2017fault}.
	We estimate that the fault-tolerant threshold for the Fibonacci topological with pair-creation noise and measurement noise lies between $p=10^{-4}$ and $p=5\cdot 10^{-4}$, which corresponds to an inverse temperature between $\beta = 9.2 /\Delta $ and $\beta = 7.6 /\Delta$. 
	This is comparable to the threshold found for the Ising topological code \cite{dauphinais2017fault}, and only one order of magnitude below that for the surface code under similar circumstances \cite{harrington}.
	For physical error rates near $p=q=10^{-4}$, corresponding to a temperature $1/\beta$ one order of magnitude below the spectral gap, a code of linear size $L=27$ yields logical error rates of the order $10^{-8}$.

	A second round of simulations was performed to determine average memory lifetimes with the assumption that all corrections happen instantaneously. 
	While this is akin to the Abelian topological codes, where distant anyons can be fused using unitary string-operators, this scenario is unphysical for non-Abelian anyons as they do not admit unitary string-operators. 
	Nevertheless, it is worth studying to which extend the results in \figref{fig:results} are influenced by the restriction to non-instantaneous recovery operations.
	In Ref.~\cite{dauphinais2017fault} it was conjectured that allowing instantaneous corrections does not significantly change the qualitative behavior of the average memory lifetimes as a function of the error rate, but mostly just increases the memory lifetimes.
	Our results, shown in  \figref{fig:results_instant}, confirm this hypothesis.

	\section{Discussion and outlook}\label{sec:discussion}
    The results presented in this work demonstrate that fault-tolerant error correction is possible for non-Abelian topological quantum error correcting codes supporting a universal logical gate set within their code space.  
    For a code consisting of Fibonacci anyons in hexagonal tiles on a two-dimensional torus, subjected to pair creation noise and measurement noise, we demonstrated that the cellular automaton decoder detailed in this work is fault-tolerant. 
    In particular, for physical error rates $p\leq10^{-3}$, it was found that the logical memory lifetime surpasses the physical coherence time for all system sizes.
    When interpreting the pair-creation noise as resulting from a non-zero temperature, this pseudo-threshold corresponds to an inverse temperature $\beta = 6.9/\Delta$, where $\Delta$ is the energy required to create a pair of Fibonacci anyons.
    Furthermore, our results suggest that this code admits a fault-tolerant quantum error correction threshold around $p = 10^{-4}$, or $\beta = 9.2 / \Delta$, which is similar to the fault-tolerant threshold found for the Ising topological code \cite{dauphinais2017fault}.

	Several future research directions present themselves.
	First, more research on a possible fault-tolerant threshold is necessary.
    Wile the numeric results presented in this work provide a strong indication that a fault-tolerant error correction threshold exists, they do not conclusively prove its existence, nor do they provide a precise estimate of its value. 
	Hence, an important open problem is the formulation of a mathematical proof of its existence. 
	Such proofs were previously formulated for the toric code \cite{harrington} and for non-cyclic non-Abelian anyon models such as in the Ising topological code \cite{dauphinais2017fault}. 
	Due to the cyclic nature of Fibonacci anyons (or any universal anyon model), however, the existing proofs are not sufficient.
	
	Second, it would be interesting to study different decoders in an identical setting. This includes both different cellular-automaton decoders such as those in Refs.~\cite{herold2015cellular, herold2017cellular}, as well as new decoders tailored to the Fibonacci topological code. 
	
	Third, while this work demonstrates that the Fibonacci topological code can be operated fault-tolerantly as a quantum memory, results regarding its use for fault-tolerant quantum computing are still lacking. 
	We envisage that fault-tolerant topological quantum computing at non-zero temperatures could be achieved by combining the code and decoding procedure presented in this work with the scheme for performing Dehn twists presented in Refs.~\cite{PhysRevLett.125.050502, PhysRevB.102.075105, Lavasani2019universal}.   Alternatively, one can also perform transversal logical gates in a folded Fibonacci code \cite{Zhu:2017tr}.
	
	Finally, it would be of great interest to expand the current results to microscopic models for non-Abelian topological quantum error correction, such as the Fibonacci Turaev-Viro code \cite{schotte2022quantum}.
	
	\section*{Acknowledgments}
	The authors would like to thank Guillaume Dauphinais and Jim Harrington for enlightening discussions on the cellular automaton decoder. 
	The computational resources (Stevin Supercomputer Infrastructure) and services used in this work were provided by the Flemish Supercomputer Center (VSC), funded by Ghent University, the Research Foundation Flanders (FWO), and the Flemish Government. 
	AS was supported by a fellowship of the Belgian American Educational Foundation. LB was supported by a PhD fellowship from the FWO. GZ was supported by the U.S. Department of Energy, Office of Science, National Quantum Information Science Research Centers, Co-design Center for Quantum Advantage (C2QA) under contract number DE-SC0012704.
	
	\appendix 
	\section{Fibonacci anyons} \label{sec:anyons}
	Below, we give with a brief overview of the topological aspects of our model. A thorough exposition of the theory of anyon models \cite{turaev1992state, kitaev2006anyons, wang2010topological, turaev2017monoidal} is beyond the scope of this work, and we restrict to a basic description of the aspects of the Fibonacci model that are required for the specific purpose of our simulations. We refer to Ref.~\cite{pfeifer2012translation} for an in-depth discussion of anyonic fusion states on the torus.
	
	The Fibonacci anyon model has two particle types, $ \1 $ and $ \tau $, that satisfy the fusion rules
	\begin{equation}\label{eq:Fib_fusion_rules}
		\1 \times \1 = \1\,, \quad \1 \times \tau = \tau \times \1 = \tau\,, \quad \tau \times \tau = \1 + \tau\,.
	\end{equation}
	It is a \emph{non-Abelian} anyon model, as the fusion of two $ \tau $-anyons can yield two distinct outcomes.
	In this case, the fusion space $ V_{ab}^c $ associated to the fusion of anyons $ a $ and $ b $ to $ c $ is one-dimensional, and is spanned by the state vector $ \ket{a, b; c} $ which we will represent graphically as
	\begin{equation}\label{eq:fusion_state}
		\ket{a, b ; c} \rightarrow \raisebox{-.7cm}{\includegraphics[scale=.42]{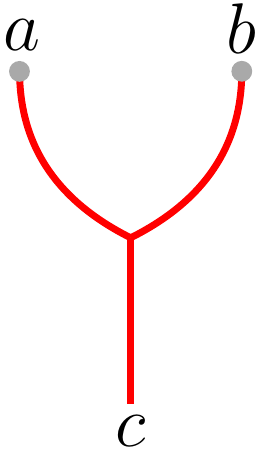}}\,.
	\end{equation}
	When fusing several anyons, their total charge is a collective property of the anyons that does not depend on the specific order in which they are fused. Mathematically, this is expressed through the associativity of the fusion rules,
	\begin{equation}\label{eq:fusion_rules_associativity}
		(a \times b) \times c = a \times (b \times c)\,.
	\end{equation}
	If we consider the case of three anyons $ a $, $ b $ and $ c $ that fuse to a total charge $ d $ then this fusion may be carried out in two distinct ways, implying the existence of two equivalent decompositions of the associated fusion space $ V_{abc}^d $ in terms of fusion states \eqref{eq:fusion_state}. These equivalent decompositions are related through a unitary transformation called an $F$-move,
	which is represented graphically as
	\begin{equation}\label{eq:F_move}
		\raisebox{-.9cm}{\includegraphics[scale=.42]{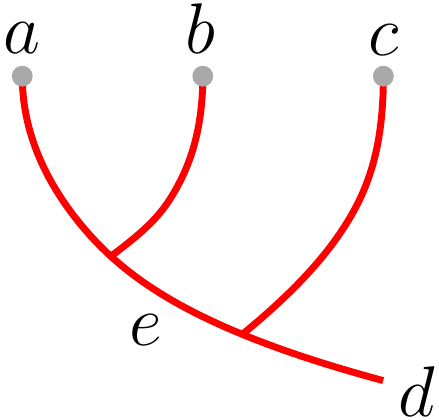}}
		\; = \sum_f F^{abe}_{cdf} \;
		\raisebox{-.9cm}{\includegraphics[scale=.42]{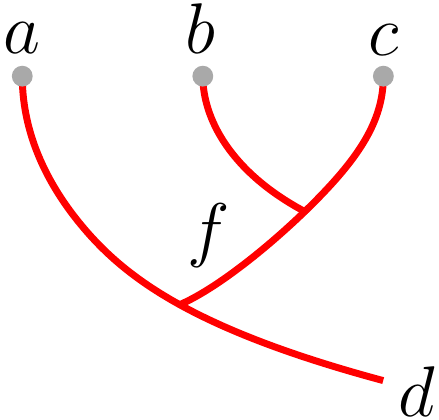}}\,.
	\end{equation}
	The coefficients $ F^{abe}_{cdf} $ in this expression
	are called \emph{$ F $-symbols} of the anyon model. 
	For the Fibonacci model they are given by
	\begin{equation}\label{eq:Fib_F_symbols_nontriv}
		F^{\tau \tau 1}_{\tau \tau 1} = \frac{1}{\phi}\,, \quad \;
		F^{\tau \tau \tau}_{\tau \tau 1} = F^{\tau \tau 1}_{\tau \tau \tau} = \frac{1}{\sqrt{\phi}}\,, \quad \;
		F^{\tau \tau \tau}_{\tau \tau \tau} = -\frac{1}{\phi}\,,
	\end{equation}
	where all other $ F $-symbols consistent with the fusion rules \eqref{eq:Fib_fusion_rules} are equal to 1, and 0 otherwise.
	
	In addition to this recoupling one can also consider the exchange or \emph{braiding} of pairs of anyons, which preserves their total charge. At the level of the fusion space, such an exchange corresponds to a basis transformation to a basis associated to a different linear ordering
	of the anyons\footnote{As opposed to the more conventional view of braiding as an active transformation that maps between different fusion spaces, we have opted for the equivalent framework of braiding as a passive basis transformation, as the latter is more appropriate in view of our specific model and numerical simulations.}. Within $ V_{ab}^c $ there are two such possible basis transformations,
	\begin{equation}\label{eq:swap}
		\raisebox{-.7cm}{\includegraphics[scale=.42]{fig/fusion_state}} =
		R^{ab}_c \raisebox{-.7cm}{\includegraphics[scale=.42]{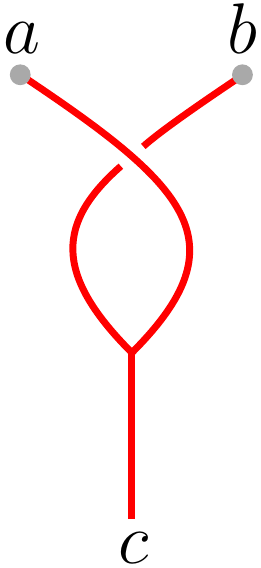}} =
		\left(R^{ba}_{c}\right)^* \raisebox{-.7cm}{\includegraphics[scale=.42]{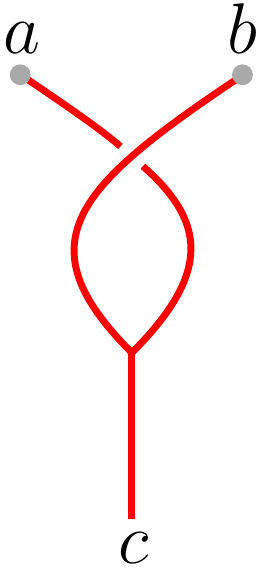}}\,,
	\end{equation}
	which we will refer to as a \emph{clockwise} and a \emph{counterclockwise swap} respectively.
	For the Fibonacci model the $R$-symbols appearing in these expressions are given by
	\begin{equation}\label{eq:Fib_R_symbols_nontriv}
		R^{\tau\tau}_1 = e^{\frac{4\pi i}{5}}\,, \qquad
		R^{\tau\tau}_\tau =  e^{-\frac{3\pi i}{5}}\,,
	\end{equation}
	where all other $ R^{ab}_c $ allowed by the fusion rules are equal to 1, and 0 otherwise.
	
\begin{figure}[H]
		\centering
		\begin{subfigure}{\linewidth}
			\centering
			\includegraphics[width=.7\linewidth]{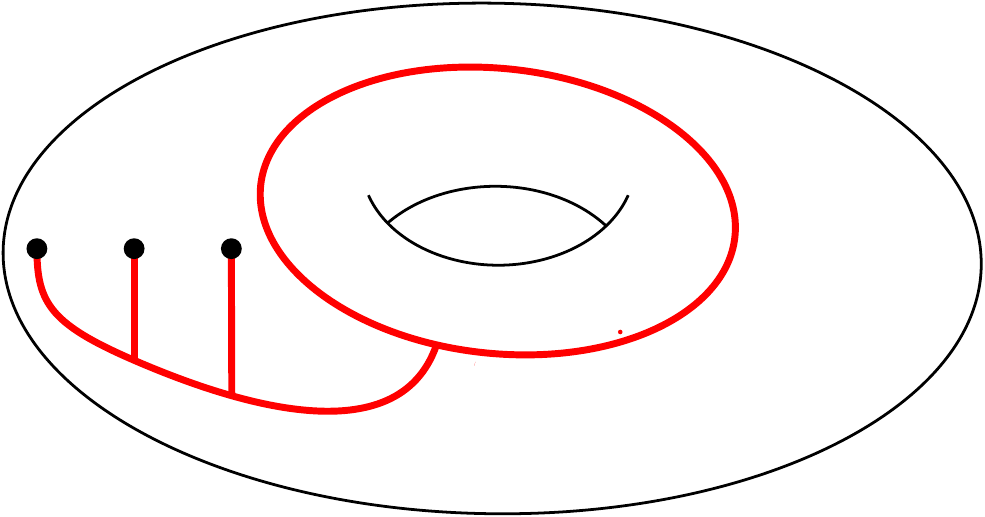}
			\caption{}
			\label{fig:inside}
		\end{subfigure}
		\begin{subfigure}{\linewidth}
			\centering
			\includegraphics[width=.7\linewidth]{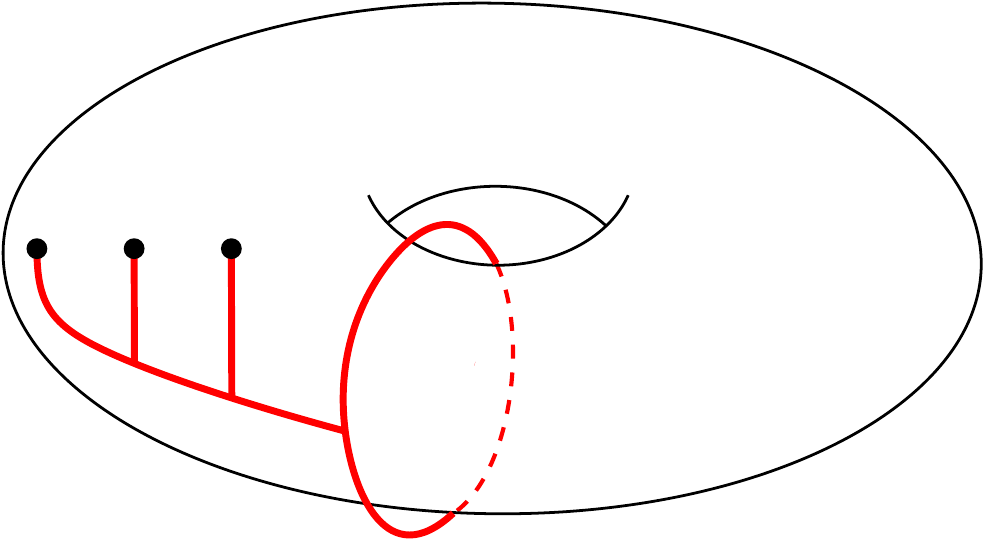}
			\caption{}
			\label{fig:outside}
		\end{subfigure}
		\caption{Two possible basis choices for anyons on a torus: a) the \emph{inside} basis, b) the \emph{outside} basis. }
		\label{fig:basis_torus}
	\end{figure}

	As we are dealing with a system that allows for an extensive amount of anyonic excitations, we will be interested in fusion states of many anyons, $ a_1, a_2, ..., a_n $, with some total charge $c$. This gives rise to an exponentially large topological Hilbert space $ V_{a_1 a_2 \cdots a_n}^c $ spanned basis states of the form
	\begin{equation}\label{eq:standard_basis_state}
		\raisebox{-.4cm}{\includegraphics[scale=.42]{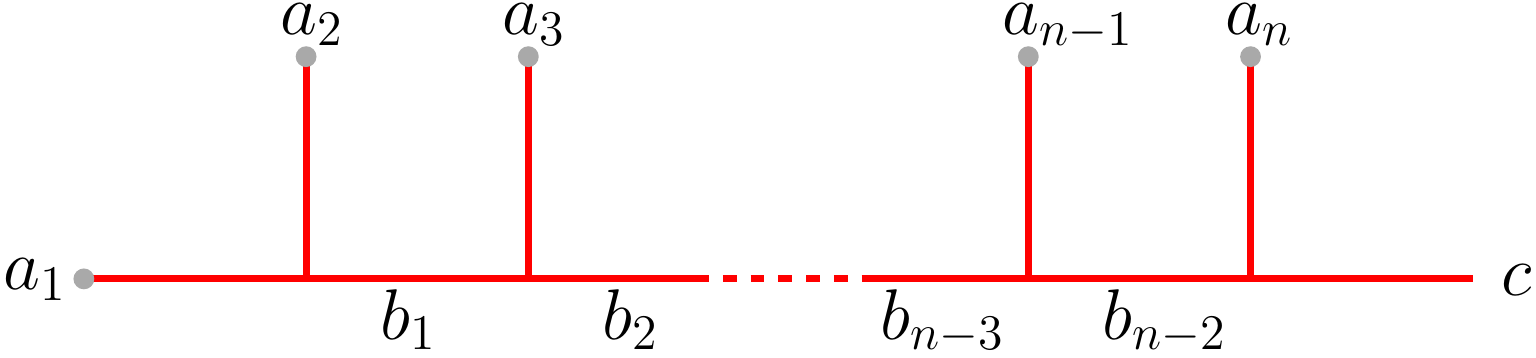}}\;.
	\end{equation}
	When dealing with anyonic fusion states on surfaces of higher genus, one must take into account additional degrees of freedom in these states that are related to the anyonic charge that runs along non-contractible cycles. 
	On a torus, this results in two distinct descriptions of fusion states, which are related through a basis change. These are known as the \emph{inside} and \emph{outside} bases, and are depicted in \figref{fig:basis_torus}.
	A detailed discussion can be found in Ref.~\cite{pfeifer2012translation}. 
	We will simply refer to these additional degrees of freedom as the \emph{handle labels} of the state. 
	
	For our current purpose, we won't need the full description of said handle labels. It is sufficient for us to pick one basis, and subsequently set the total charge of all anyons to the vacuum. We are left with only a single label ($\1$ or $\tau$) representing the anyonic charge flowing along the non-contractible cycle associated with our basis choice. This is precisely the origin of the twofold degeneracy of the anyonic vacuum on the torus, which we have taken to be our code space.
	
	In the this work we will always start from the code state corresponding to a trivial handle label. As a change in handle label at any point during the error correction procedure must be the consequence of a topologically non-trivial process which constitutes a logical error, any simulation is aborted at the occurrence of such an event, meaning that all handle labels can be safely ignored in the remainder of our discussion.
	
	A basic functionality required for the simulation of the error correction procedure is the ability to correctly sample a measurement of the total charge of a given set of anyons. Starting from a given fusion state, this can be achieved by first transforming to a basis in which the relevant anyons are fused sequentially. For a system of many anyons these reordering basis transformations are obtained by combining Eqs.~\eqref{eq:F_move} and \eqref{eq:swap}, giving rise to the clockwise swap
	\begin{multline}\label{eq:swap_cw_comp}
		\raisebox{-.5cm}{\includegraphics[scale=.38]{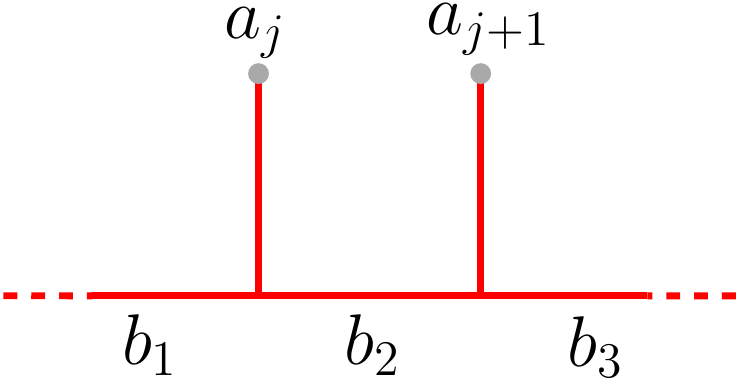}} \\
		= 
		\sum_{b_2'} B^{b_1 a_{j} b_2}_{a_{j+1} b_3 b_2'} \;\;
		\raisebox{-.5cm}{\includegraphics[scale=.38]{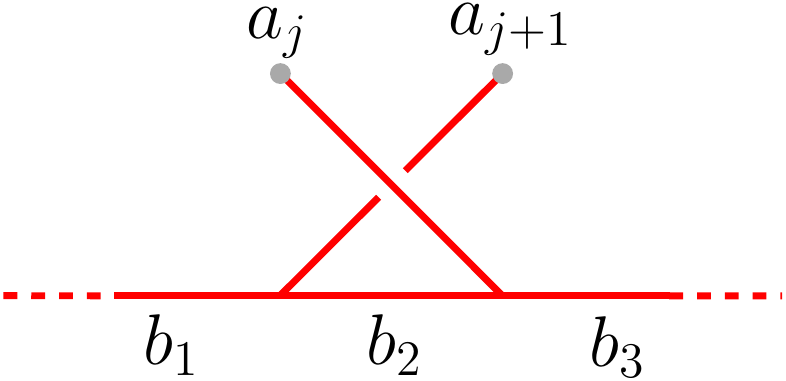}}\;, 
	\end{multline}
	and the counterclockwise swap
	\begin{multline}\label{eq:swap_ccw_comp}
		\raisebox{-.5cm}{\includegraphics[scale=.38]{fig/cw_braid_diagram_initial.pdf}} \\
		=
		\sum_{b_2'} \left(B^{b_1 a_{j} b_2}_{a_{j+1} b_3 b_2'} \right)^* \;
		\raisebox{-.5cm}{\includegraphics[scale=.38]{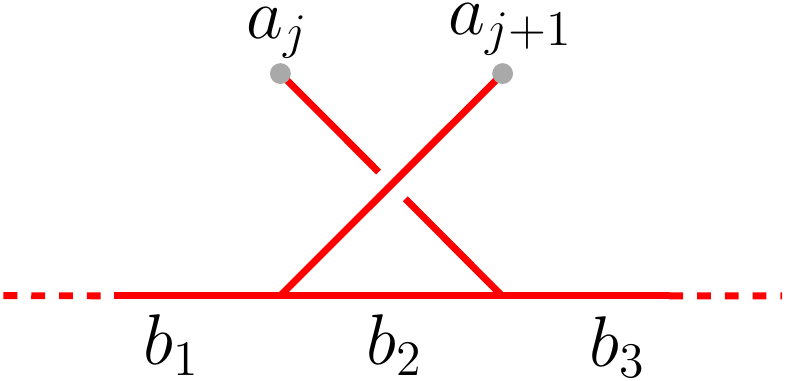}} \,,
	\end{multline}
	where 
	\begin{equation}\label{eq:B_symbol}
		B^{b_1 a_{j} b_2}_{a_{j+1} b_3 b_2'} = \sum_{c} 
		F^{a_{j+1}  a_{j} c}_{ b_3  b_1  b_2'} 
		R^{a_{j} a_{j+1}}_{c} 
		F^{b_1 a_{j} b_2}_{a_{j+1} b_3 c}\;.
	\end{equation}
	By performing a certain set of these basis transformations, the fusion order can always be made consistent with the group of anyons of which we want to measure the total charge. Subsequently, the fusion state is recoupled such that the relevant group of anyons is connected to the rest of the state by a single edge $c$. For a charge measurement of a pair of anyons this recoupling takes the form
	\begin{equation}\label{eq:F_move_comp}
		\raisebox{-.5cm}{\includegraphics[scale=.38]{fig/cw_braid_diagram_initial.pdf}} = 
		\sum_{c} F^{b_1 a_{j} b_2}_{a_{j+1} b_3 c} \;
		\raisebox{-.5cm}{\includegraphics[scale=.38]{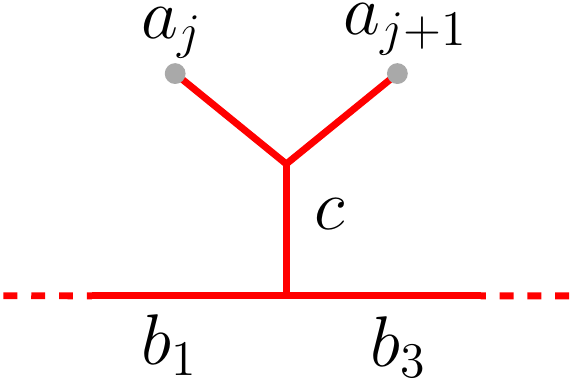}} \,,
	\end{equation}
	and charge measurements of larger groups of anyons simply require consecutive applications of the recoupling identity \eqref{eq:F_move}. Finally, the charge outcome is obtained sampling the total charge $c$ from the probability distribution corresponding to the resulting superposition of fusion states.

	\section{Classical simulatbility}
\begin{figure}
		\begin{subfigure}{.8\linewidth}
			\centering
			\includegraphics[scale=0.5]{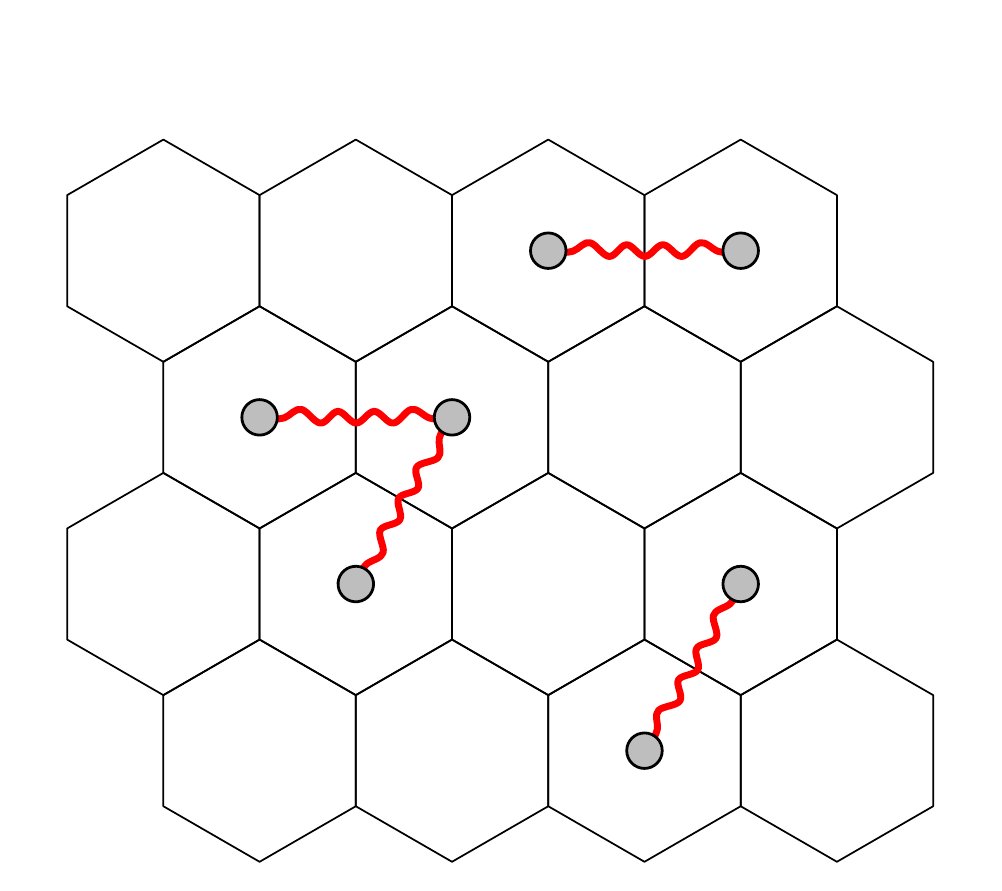}
			\caption{}
		\end{subfigure}
		\begin{subfigure}{.8\linewidth}
			\centering
			\includegraphics[scale=0.5]{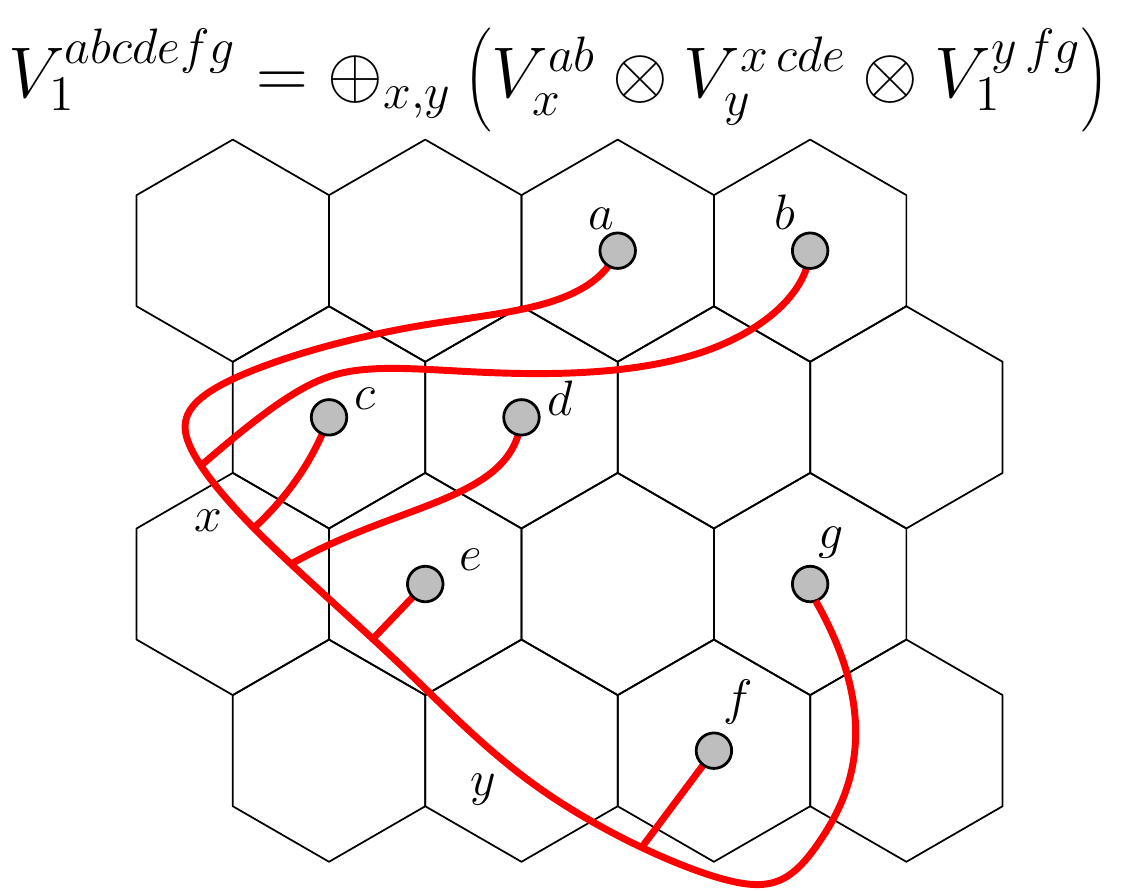}
			\caption{}
		\end{subfigure}
		\begin{subfigure}{.8\linewidth}
			\centering
			\includegraphics[scale=0.5]{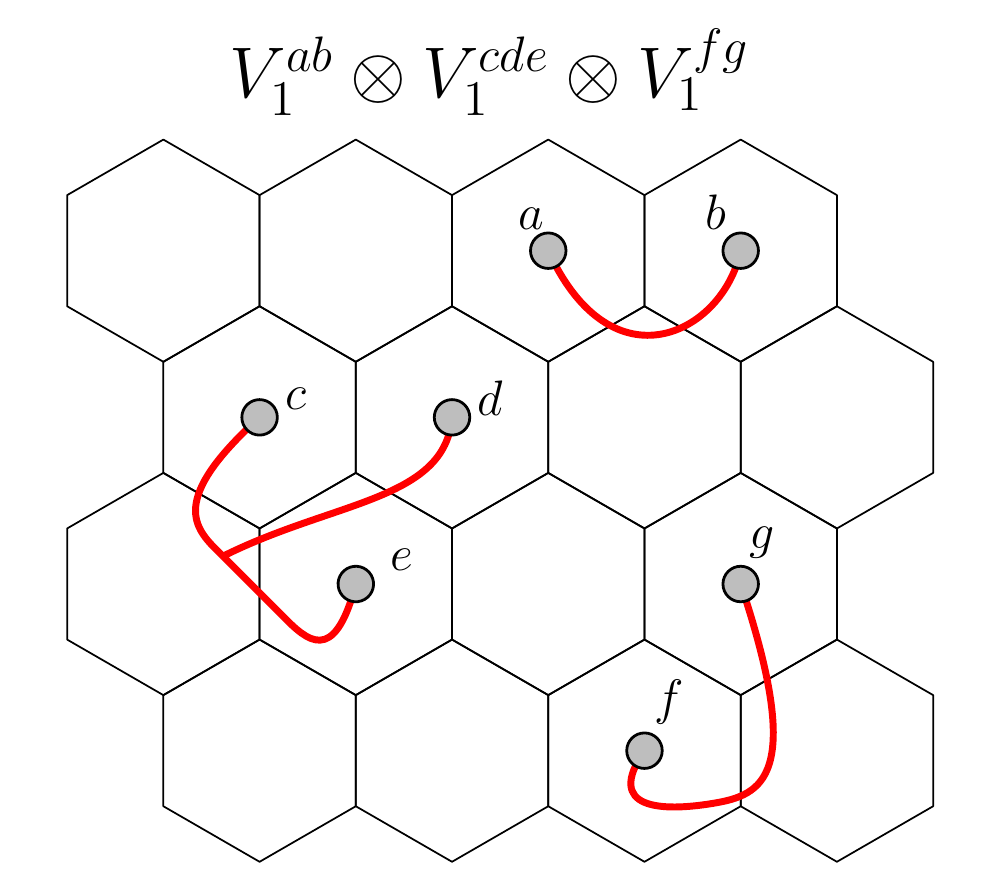}
			\caption{}
		\end{subfigure}
		\caption{(a) A set of noise processes. (b) A generic state in the full fusion space of all anyons created by the noise processes involves superpositions in the labels $x$ and $y$. (c) The factorized Hilbert space which is sufficient to represent the state.}
		\label{fig:fusion_space}
	\end{figure}

	It is well known that Fibonacci anyons are universal for quantum computation \cite{freedman2002modular}. 
	One might therefore be tempted to conclude that the classical simulation of the topological Fibonacci code described in Sec.~\ref{sec:code} with pair-creation noise is unlikely to succeed.
	However, as was noted in Ref.~\cite{burton2017classical}, the simulation of noise and error correction processes does not require the simulation of \emph{general} anyon dynamics. 
	In particular, 	individual noise processes create distinct connected groups of anyons with vacuum total charge (or extend such existing groups). These groups correspond to anyons that have interacted at some point during their lifetime, and must thus only be merged whenever a noise or error correction process involves two members from disconnected groups. 
	Since each connected group has a trivial total charge, braiding between disconnected groups is trivial.
	Hence, the total fusion space factorizes into a tensor product of fusion spaces of individual connected groups, and we are only required to simulate anyon dynamics within each of these groups separately. This factorization of the fusion space is illustrated in \figref{fig:fusion_space}.
	
	The creation and subsequent merging of disconnected groups of anyons by noise and recovery processes can be thought of as a kind of percolation process. 
	Hence, below the percolation threshold, one expects that the size of the largest connected group scales as $ O(\log(L)) $ (with variance $ O(1) $), where $ L $ is the linear system size \cite{bazant2000largest}.
	As this is a probabilistic statement, there will be instances where the largest connected group has a size larger than $ O(\log(L) $, but the probability of such events is suppressed exponentially with the system size $ L $. 
	This logarithmic scaling of the largest cluster size $ s = O(\log(L)) $ counters the exponential scaling of the dimension of the fusion space $ d = O(\exp(s)) $ for individual connected groups. 
	Therefore, the fusion spaces of individual connect groups will have dimension $  \text{dim} = O(\text{poly}(L)) $, meaning that the dynamics within connected groups can be simulated efficiently.
	
	Exploiting the tensor product structure within the total fusion space, requires the use of basis for the anyonic fusion space which reflects this structure and can be updated dynamically to keep track of noise and recovery processes. 
	This is best achieved by using the framework of \emph{curve diagrams}, which were introduced in Ref.~\cite{burton2017classical} (also see Ref.~\cite{burton2016short} for a more rigorous treatment using the language of modular functors), and also discussed extensively in Ref.~\cite{schotte2022quantum}. Since these are merely a technical tool for keeping track of the most appropriate basis during the numerical simulations, we will refrain from discussing them here. Interested readers are referred to the aforementioned references for details.

\section{Transition rules for $ Q = 3 $} \label{sec:transition_rules}
	Below, we give a full definition of the transition rules for $ Q = 3 $. 
	It is possible to define more general transition rules that apply for any colony size, examples of such rules can be found in Refs.~\cite{harrington} and \cite{dauphinais2017fault}. However, since the numerical simulations performed in this work were performed to $ Q=3 $ we have taken the freedom to tailor the transition rules to this case specifically.


	\begin{itemize}
		\item North-West\\
		if $ s_{k,c}(\rho) = 0$, do nothing;\\
		else if $ s_{k,n}(\rho + (0,-1)) \neq 0$, do nothing; \\	
		else if $ s_{k,n}(\rho + (-1,0)) \neq 0$, do nothing; \\	
		else if $ s_{k,n}(\rho + (0,1)) \neq 0$, move east; \\		
		else if $ s_{k,n}(\rho + (1,0)) \neq 0$, move south; \\	
		else if  $ s_{k,n}(\rho + (1,-1)) \neq 0$, do nothing; \\ 	
		else if  $ s_{k,n}(\rho + (-1,-1)) \neq 0$, do nothing; \\	
		else if  $ s_{k,n}(\rho + (-1,1)) \neq 0$, do nothing;	\\ 	
		else move south;
		
		\item North\\
		if $ s_{k,c}(\rho) = 0$, do nothing;\\
		else if $ s_{k,n}(\rho + (-1,0)) \neq 0$, do nothing; \\	
		else if $ s_{k,n}(\rho + (0,-1)) \neq 0$, do nothing; \\	
		else if $ s_{k,n}(\rho + (0,1)) \neq 0$, do nothing; \\	
		else if  $ s_{k,n}(\rho + (-1,1)) \neq 0$, do nothing;	\\ 	
		else if  $ s_{k,n}(\rho + (-1,-1)) \neq 0$, do nothing; \\	
		else move south;
		
		\item North-East\\
		if $ s_{k,c}(\rho) = 0$, do nothing;\\
		else if $ s_{k,n}(\rho + (0,1)) \neq 0$, move east; \\		
		else if $ s_{k,n}(\rho + (-1,1)) \neq 0$, move north-east; \\	
		else if $ s_{k,n}(\rho + (1,1)) \neq 0$, move south; \\		
		else if $ s_{k,n}(\rho + (-1,0)) \neq 0$, do nothing; \\	
		else if $ s_{k,n}(\rho + (0,-1)) \neq 0$, move west; \\		
		else if $ s_{k,n}(\rho + (1,0)) \neq 0$, move south; \\		
		else if  $ s_{k,n}(\rho + (-1,-1)) \neq 0$, do nothing; \\	
		else move south-west; \vspace{5pt}
		
		\item East\\
		if $ s_{k,c}(\rho) = 0$, do nothing;\\
		else if $ s_{k,n}(\rho + (0,1)) \neq 0$, move east; \\		
		else if $ s_{k,n}(\rho + (-1,1)) \neq 0$, move north-east; \\	
		else if $ s_{k,n}(\rho + (1,1)) \neq 0$, move south; \\	
		else if $ s_{k,n}(\rho + (-1,0)) \neq 0$, do nothing; \\	
		else if $ s_{k,n}(\rho + (1,0)) \neq 0$, do nothing; \\		
		else move west;
		
		\item South-East\\
		if $ s_{k,c}(\rho) = 0$, do nothing;\\
		else if $ s_{k,n}(\rho + (0,1)) \neq 0$, move east; \\		
		else if $ s_{k,n}(\rho + (1,0)) \neq 0$, move south; \\	
		else if $ s_{k,n}(\rho + (-1,1)) \neq 0$, move north-east; \\	
		else if  $ s_{k,n}(\rho + (1,-1)) \neq 0$, move south-west; \\ 
		else if $ s_{k,n}(\rho + (1,1)) \neq 0$, move south; \\	
		else if $ s_{k,n}(\rho + (0,-1)) \neq 0$, move west; \\		
		else move north;							
		
		\item South\\
		if $ s_{k,c}(\rho) = 0$, do nothing;\\
		else if $ s_{k,n}(\rho + (1,0)) \neq 0$, move south; \\	
		else if  $ s_{k,n}(\rho + (1,-1)) \neq 0$, move south-west; \\ 
		else if  $ s_{k,n}(\rho + (1,1)) \neq 0$, move east; \\ 	
		else if $ s_{k,n}(\rho + (0,-1)) \neq 0$, do nothing; \\	
		else if $ s_{k,n}(\rho + (0,1)) \neq 0$, do nothing; \\	
		else move north;
		
		\item South-West\\
		if $ s_{k,c}(\rho) = 0$, do nothing;\\
		else if $ s_{k,n}(\rho + (1,0)) \neq 0$, move south; \\	
		else if  $ s_{k,n}(\rho + (1,-1)) \neq 0$, move south-west; \\ 
		else if  $ s_{k,n}(\rho + (1,1)) \neq 0$, move east; \\ 	
		else if $ s_{k,n}(\rho + (0,-1)) \neq 0$, do nothing; \\	
		else if $ s_{k,n}(\rho + (-1,0)) \neq 0$, move north; \\	
		else if $ s_{k,n}(\rho + (0,1)) \neq 0$, move east; \\	
		else if  $ s_{k,n}(\rho + (-1,-1)) \neq 0$, do nothing; \\	
		else move north-east;				
		
		\item West\\
		if $ s_{k,c}(\rho) = 0$, do nothing;\\
		else if $ s_{k,n}(\rho + (0,-1)) \neq 0$, do nothing; \\	
		else if $ s_{k,n}(\rho + (-1,0)) \neq 0$, do nothing; \\	
		else if $ s_{k,n}(\rho + (1,0)) \neq 0$, do nothing; \\		
		else if  $ s_{k,n}(\rho + (1,-1)) \neq 0$, do nothing; \\ 	
		else if  $ s_{k,n}(\rho + (-1,-1)) \neq 0$, do nothing; \\	
		else move east;
		
	\end{itemize}
	
	\bibliography{references, mybib_merge}

\begin{thebibliography}{44}%
\makeatletter
\providecommand \@ifxundefined [1]{%
 \@ifx{#1\undefined}
}%
\providecommand \@ifnum [1]{%
 \ifnum #1\expandafter \@firstoftwo
 \else \expandafter \@secondoftwo
 \fi
}%
\providecommand \@ifx [1]{%
 \ifx #1\expandafter \@firstoftwo
 \else \expandafter \@secondoftwo
 \fi
}%
\providecommand \natexlab [1]{#1}%
\providecommand \enquote  [1]{``#1''}%
\providecommand \bibnamefont  [1]{#1}%
\providecommand \bibfnamefont [1]{#1}%
\providecommand \citenamefont [1]{#1}%
\providecommand \href@noop [0]{\@secondoftwo}%
\providecommand \href [0]{\begingroup \@sanitize@url \@href}%
\providecommand \@href[1]{\@@startlink{#1}\@@href}%
\providecommand \@@href[1]{\endgroup#1\@@endlink}%
\providecommand \@sanitize@url [0]{\catcode `\\12\catcode `\$12\catcode
  `\&12\catcode `\#12\catcode `\^12\catcode `\_12\catcode `\%12\relax}%
\providecommand \@@startlink[1]{}%
\providecommand \@@endlink[0]{}%
\providecommand \url  [0]{\begingroup\@sanitize@url \@url }%
\providecommand \@url [1]{\endgroup\@href {#1}{\urlprefix }}%
\providecommand \urlprefix  [0]{URL }%
\providecommand \Eprint [0]{\href }%
\providecommand \doibase [0]{https://doi.org/}%
\providecommand \selectlanguage [0]{\@gobble}%
\providecommand \bibinfo  [0]{\@secondoftwo}%
\providecommand \bibfield  [0]{\@secondoftwo}%
\providecommand \translation [1]{[#1]}%
\providecommand \BibitemOpen [0]{}%
\providecommand \bibitemStop [0]{}%
\providecommand \bibitemNoStop [0]{.\EOS\space}%
\providecommand \EOS [0]{\spacefactor3000\relax}%
\providecommand \BibitemShut  [1]{\csname bibitem#1\endcsname}%
\let\auto@bib@innerbib\@empty
\bibitem [{\citenamefont {G{\'a}cs}(1986)}]{gacs1986reliable}%
  \BibitemOpen
  \bibfield  {author} {\bibinfo {author} {\bibfnamefont {P.}~\bibnamefont
  {G{\'a}cs}},\ }\bibfield  {title} {\bibinfo {title} {Reliable computation
  with cellular automata},\ }\href
  {https://doi.org/https://doi.org/10.1016/0022-0000(86)90002-4} {\bibfield
  {journal} {\bibinfo  {journal} {Journal of Computer and System Sciences}\
  }\textbf {\bibinfo {volume} {32}},\ \bibinfo {pages} {15} (\bibinfo {year}
  {1986})}\BibitemShut {NoStop}%
\bibitem [{\citenamefont {Harrington}(2004)}]{harrington}%
  \BibitemOpen
  \bibfield  {author} {\bibinfo {author} {\bibfnamefont {J.~W.}\ \bibnamefont
  {Harrington}},\ }\emph {\bibinfo {title} {Analysis of quantum
  error-correcting codes: symplectic lattice codes and toric codes}},\ \href
  {https://doi.org/https://doi.org/10.7907/AHMQ-EG82} {Ph.D. thesis},\ \bibinfo
   {school} {California Institute of Technology} (\bibinfo {year}
  {2004})\BibitemShut {NoStop}%
\bibitem [{\citenamefont {Freedman}\ \emph
  {et~al.}(2002{\natexlab{a}})\citenamefont {Freedman}, \citenamefont
  {Larsen},\ and\ \citenamefont {Wang}}]{freedman2002modular}%
  \BibitemOpen
  \bibfield  {author} {\bibinfo {author} {\bibfnamefont {M.~H.}\ \bibnamefont
  {Freedman}}, \bibinfo {author} {\bibfnamefont {M.}~\bibnamefont {Larsen}},\
  and\ \bibinfo {author} {\bibfnamefont {Z.}~\bibnamefont {Wang}},\ }\bibfield
  {title} {\bibinfo {title} {A {{Modular Functor Which}} is {{Universal}} for
  {{Quantum Computation}}},\ }\href {https://doi.org/10.1007/s002200200645}
  {\bibfield  {journal} {\bibinfo  {journal} {Communications in Mathematical
  Physics}\ }\textbf {\bibinfo {volume} {227}},\ \bibinfo {pages} {605}
  (\bibinfo {year} {2002}{\natexlab{a}})}\BibitemShut {NoStop}%
\bibitem [{\citenamefont {Freedman}\ \emph
  {et~al.}(2002{\natexlab{b}})\citenamefont {Freedman}, \citenamefont
  {Kitaev},\ and\ \citenamefont {Wang}}]{freedman2002simulation}%
  \BibitemOpen
  \bibfield  {author} {\bibinfo {author} {\bibfnamefont {M.~H.}\ \bibnamefont
  {Freedman}}, \bibinfo {author} {\bibfnamefont {A.}~\bibnamefont {Kitaev}},\
  and\ \bibinfo {author} {\bibfnamefont {Z.}~\bibnamefont {Wang}},\ }\bibfield
  {title} {\bibinfo {title} {Simulation of topological field theories by
  quantum computers},\ }\href@noop {} {\bibfield  {journal} {\bibinfo
  {journal} {Communications in Mathematical Physics}\ }\textbf {\bibinfo
  {volume} {227}},\ \bibinfo {pages} {587} (\bibinfo {year}
  {2002}{\natexlab{b}})}\BibitemShut {NoStop}%
\bibitem [{\citenamefont {Kitaev}(2003)}]{kitaev2003fault}%
  \BibitemOpen
  \bibfield  {author} {\bibinfo {author} {\bibfnamefont {A.}~\bibnamefont
  {Kitaev}},\ }\bibfield  {title} {\bibinfo {title} {Fault-tolerant quantum
  computation by anyons},\ }\href
  {https://doi.org/https://doi.org/10.1016/S0003-4916(02)00018-0} {\bibfield
  {journal} {\bibinfo  {journal} {Annals of Physics}\ }\textbf {\bibinfo
  {volume} {303}},\ \bibinfo {pages} {2} (\bibinfo {year} {2003})}\BibitemShut
  {NoStop}%
\bibitem [{\citenamefont {Bravyi}\ and\ \citenamefont
  {Kitaev}(1998)}]{bravyi1998quantum}%
  \BibitemOpen
  \bibfield  {author} {\bibinfo {author} {\bibfnamefont {S.~B.}\ \bibnamefont
  {Bravyi}}\ and\ \bibinfo {author} {\bibfnamefont {A.~Y.}\ \bibnamefont
  {Kitaev}},\ }\bibfield  {title} {\bibinfo {title} {Quantum codes on a lattice
  with boundary},\ }\href@noop {} {\bibfield  {journal} {\bibinfo  {journal}
  {arXiv preprint quant-ph/9811052}\ } (\bibinfo {year} {1998})}\BibitemShut
  {NoStop}%
\bibitem [{\citenamefont {Dennis}\ \emph {et~al.}(2002)\citenamefont {Dennis},
  \citenamefont {Kitaev}, \citenamefont {Landahl},\ and\ \citenamefont
  {Preskill}}]{dennis2002topological}%
  \BibitemOpen
  \bibfield  {author} {\bibinfo {author} {\bibfnamefont {E.}~\bibnamefont
  {Dennis}}, \bibinfo {author} {\bibfnamefont {A.}~\bibnamefont {Kitaev}},
  \bibinfo {author} {\bibfnamefont {A.}~\bibnamefont {Landahl}},\ and\ \bibinfo
  {author} {\bibfnamefont {J.}~\bibnamefont {Preskill}},\ }\bibfield  {title}
  {\bibinfo {title} {Topological quantum memory},\ }\href
  {https://doi.org/10.1063/1.1499754} {\bibfield  {journal} {\bibinfo
  {journal} {Journal of Mathematical Physics}\ }\textbf {\bibinfo {volume}
  {43}},\ \bibinfo {pages} {4452} (\bibinfo {year} {2002})}\BibitemShut
  {NoStop}%
\bibitem [{\citenamefont {Wang}\ \emph {et~al.}(2009)\citenamefont {Wang},
  \citenamefont {Fowler}, \citenamefont {Stephens},\ and\ \citenamefont
  {Hollenberg}}]{wang2009threshold}%
  \BibitemOpen
  \bibfield  {author} {\bibinfo {author} {\bibfnamefont {D.~S.}\ \bibnamefont
  {Wang}}, \bibinfo {author} {\bibfnamefont {A.~G.}\ \bibnamefont {Fowler}},
  \bibinfo {author} {\bibfnamefont {A.~M.}\ \bibnamefont {Stephens}},\ and\
  \bibinfo {author} {\bibfnamefont {L.~C.}\ \bibnamefont {Hollenberg}},\
  }\bibfield  {title} {\bibinfo {title} {Threshold error rates for the toric
  and surface codes},\ }\href@noop {} {\bibfield  {journal} {\bibinfo
  {journal} {arXiv preprint arXiv:0905.0531}\ } (\bibinfo {year}
  {2009})}\BibitemShut {NoStop}%
\bibitem [{\citenamefont {Fowler}\ \emph {et~al.}(2012)\citenamefont {Fowler},
  \citenamefont {Mariantoni}, \citenamefont {Martinis},\ and\ \citenamefont
  {Cleland}}]{fowler2012surface}%
  \BibitemOpen
  \bibfield  {author} {\bibinfo {author} {\bibfnamefont {A.~G.}\ \bibnamefont
  {Fowler}}, \bibinfo {author} {\bibfnamefont {M.}~\bibnamefont {Mariantoni}},
  \bibinfo {author} {\bibfnamefont {J.~M.}\ \bibnamefont {Martinis}},\ and\
  \bibinfo {author} {\bibfnamefont {A.~N.}\ \bibnamefont {Cleland}},\
  }\bibfield  {title} {\bibinfo {title} {Surface codes: Towards practical
  large-scale quantum computation},\ }\href
  {https://doi.org/10.1103/PhysRevA.86.032324} {\bibfield  {journal} {\bibinfo
  {journal} {Phys. Rev. A}\ }\textbf {\bibinfo {volume} {86}},\ \bibinfo
  {pages} {032324} (\bibinfo {year} {2012})}\BibitemShut {NoStop}%
\bibitem [{\citenamefont {Wootton}\ and\ \citenamefont
  {Loss}(2012)}]{wootton2012high}%
  \BibitemOpen
  \bibfield  {author} {\bibinfo {author} {\bibfnamefont {J.~R.}\ \bibnamefont
  {Wootton}}\ and\ \bibinfo {author} {\bibfnamefont {D.}~\bibnamefont {Loss}},\
  }\bibfield  {title} {\bibinfo {title} {High threshold error correction for
  the surface code},\ }\href@noop {} {\bibfield  {journal} {\bibinfo  {journal}
  {Physical review letters}\ }\textbf {\bibinfo {volume} {109}},\ \bibinfo
  {pages} {160503} (\bibinfo {year} {2012})}\BibitemShut {NoStop}%
\bibitem [{\citenamefont {Anwar}\ \emph {et~al.}(2014)\citenamefont {Anwar},
  \citenamefont {Brown}, \citenamefont {Campbell},\ and\ \citenamefont
  {Browne}}]{anwar2014fast}%
  \BibitemOpen
  \bibfield  {author} {\bibinfo {author} {\bibfnamefont {H.}~\bibnamefont
  {Anwar}}, \bibinfo {author} {\bibfnamefont {B.~J.}\ \bibnamefont {Brown}},
  \bibinfo {author} {\bibfnamefont {E.~T.}\ \bibnamefont {Campbell}},\ and\
  \bibinfo {author} {\bibfnamefont {D.~E.}\ \bibnamefont {Browne}},\ }\bibfield
   {title} {\bibinfo {title} {Fast decoders for qudit topological codes},\
  }\href@noop {} {\bibfield  {journal} {\bibinfo  {journal} {New Journal of
  Physics}\ }\textbf {\bibinfo {volume} {16}},\ \bibinfo {pages} {063038}
  (\bibinfo {year} {2014})}\BibitemShut {NoStop}%
\bibitem [{\citenamefont {Bravyi}\ \emph {et~al.}(2014)\citenamefont {Bravyi},
  \citenamefont {Suchara},\ and\ \citenamefont {Vargo}}]{bravyi2014efficient}%
  \BibitemOpen
  \bibfield  {author} {\bibinfo {author} {\bibfnamefont {S.}~\bibnamefont
  {Bravyi}}, \bibinfo {author} {\bibfnamefont {M.}~\bibnamefont {Suchara}},\
  and\ \bibinfo {author} {\bibfnamefont {A.}~\bibnamefont {Vargo}},\ }\bibfield
   {title} {\bibinfo {title} {Efficient algorithms for maximum likelihood
  decoding in the surface code},\ }\href@noop {} {\bibfield  {journal}
  {\bibinfo  {journal} {Physical Review A}\ }\textbf {\bibinfo {volume} {90}},\
  \bibinfo {pages} {032326} (\bibinfo {year} {2014})}\BibitemShut {NoStop}%
\bibitem [{\citenamefont {Wootton}(2015)}]{wootton2015simple}%
  \BibitemOpen
  \bibfield  {author} {\bibinfo {author} {\bibfnamefont {J.}~\bibnamefont
  {Wootton}},\ }\bibfield  {title} {\bibinfo {title} {A simple decoder for
  topological codes},\ }\href@noop {} {\bibfield  {journal} {\bibinfo
  {journal} {Entropy}\ }\textbf {\bibinfo {volume} {17}},\ \bibinfo {pages}
  {1946} (\bibinfo {year} {2015})}\BibitemShut {NoStop}%
\bibitem [{\citenamefont {Andrist}\ \emph {et~al.}(2015)\citenamefont
  {Andrist}, \citenamefont {Wootton},\ and\ \citenamefont
  {Katzgraber}}]{andrist2015error}%
  \BibitemOpen
  \bibfield  {author} {\bibinfo {author} {\bibfnamefont {R.~S.}\ \bibnamefont
  {Andrist}}, \bibinfo {author} {\bibfnamefont {J.~R.}\ \bibnamefont
  {Wootton}},\ and\ \bibinfo {author} {\bibfnamefont {H.~G.}\ \bibnamefont
  {Katzgraber}},\ }\bibfield  {title} {\bibinfo {title} {Error thresholds for
  abelian quantum double models: Increasing the bit-flip stability of
  topological quantum memory},\ }\href@noop {} {\bibfield  {journal} {\bibinfo
  {journal} {Physical Review A}\ }\textbf {\bibinfo {volume} {91}},\ \bibinfo
  {pages} {042331} (\bibinfo {year} {2015})}\BibitemShut {NoStop}%
\bibitem [{\citenamefont {Gottesman}(1997)}]{gottesman1997stabilizer}%
  \BibitemOpen
  \bibfield  {author} {\bibinfo {author} {\bibfnamefont {D.~E.}\ \bibnamefont
  {Gottesman}},\ }\emph {\bibinfo {title} {Stabilizer {{Codes}} and {{Quantum
  Error Correction}}}},\ \href {https://doi.org/10.7907/rzr7-dt72} {Ph.D.
  thesis},\ \bibinfo  {school} {California Institute of Technology} (\bibinfo
  {year} {1997})\BibitemShut {NoStop}%
\bibitem [{\citenamefont {Bravyi}\ and\ \citenamefont
  {Kitaev}(2005)}]{bravyi2005universal}%
  \BibitemOpen
  \bibfield  {author} {\bibinfo {author} {\bibfnamefont {S.}~\bibnamefont
  {Bravyi}}\ and\ \bibinfo {author} {\bibfnamefont {A.}~\bibnamefont
  {Kitaev}},\ }\bibfield  {title} {\bibinfo {title} {Universal quantum
  computation with ideal clifford gates and noisy ancillas},\ }\href@noop {}
  {\bibfield  {journal} {\bibinfo  {journal} {Physical Review A}\ }\textbf
  {\bibinfo {volume} {71}},\ \bibinfo {pages} {022316} (\bibinfo {year}
  {2005})}\BibitemShut {NoStop}%
\bibitem [{\citenamefont {Kubica}\ and\ \citenamefont
  {Beverland}(2015)}]{kubica2015universal}%
  \BibitemOpen
  \bibfield  {author} {\bibinfo {author} {\bibfnamefont {A.}~\bibnamefont
  {Kubica}}\ and\ \bibinfo {author} {\bibfnamefont {M.~E.}\ \bibnamefont
  {Beverland}},\ }\bibfield  {title} {\bibinfo {title} {Universal transversal
  gates with color codes: A simplified approach},\ }\href@noop {} {\bibfield
  {journal} {\bibinfo  {journal} {Physical Review A}\ }\textbf {\bibinfo
  {volume} {91}},\ \bibinfo {pages} {032330} (\bibinfo {year}
  {2015})}\BibitemShut {NoStop}%
\bibitem [{\citenamefont {Bomb{\'\i}n}(2016)}]{bombin2016dimensional}%
  \BibitemOpen
  \bibfield  {author} {\bibinfo {author} {\bibfnamefont {H.}~\bibnamefont
  {Bomb{\'\i}n}},\ }\bibfield  {title} {\bibinfo {title} {Dimensional jump in
  quantum error correction},\ }\href@noop {} {\bibfield  {journal} {\bibinfo
  {journal} {New Journal of Physics}\ }\textbf {\bibinfo {volume} {18}},\
  \bibinfo {pages} {043038} (\bibinfo {year} {2016})}\BibitemShut {NoStop}%
\bibitem [{\citenamefont {Campbell}\ \emph {et~al.}(2017)\citenamefont
  {Campbell}, \citenamefont {Terhal},\ and\ \citenamefont
  {Vuillot}}]{campbell2017roads}%
  \BibitemOpen
  \bibfield  {author} {\bibinfo {author} {\bibfnamefont {E.~T.}\ \bibnamefont
  {Campbell}}, \bibinfo {author} {\bibfnamefont {B.~M.}\ \bibnamefont
  {Terhal}},\ and\ \bibinfo {author} {\bibfnamefont {C.}~\bibnamefont
  {Vuillot}},\ }\bibfield  {title} {\bibinfo {title} {Roads towards
  fault-tolerant universal quantum computation},\ }\href
  {https://doi.org/10.1038/nature23460} {\bibfield  {journal} {\bibinfo
  {journal} {Nature}\ }\textbf {\bibinfo {volume} {549}},\ \bibinfo {pages}
  {172} (\bibinfo {year} {2017})},\ \Eprint {https://arxiv.org/abs/1612.07330}
  {1612.07330} \BibitemShut {NoStop}%
\bibitem [{\citenamefont {Hutter}\ and\ \citenamefont
  {Wootton}(2016)}]{hutter2016continuous}%
  \BibitemOpen
  \bibfield  {author} {\bibinfo {author} {\bibfnamefont {A.}~\bibnamefont
  {Hutter}}\ and\ \bibinfo {author} {\bibfnamefont {J.~R.}\ \bibnamefont
  {Wootton}},\ }\bibfield  {title} {\bibinfo {title} {Continuous error
  correction for ising anyons},\ }\href@noop {} {\bibfield  {journal} {\bibinfo
   {journal} {Physical Review A}\ }\textbf {\bibinfo {volume} {93}},\ \bibinfo
  {pages} {042327} (\bibinfo {year} {2016})}\BibitemShut {NoStop}%
\bibitem [{\citenamefont {Wootton}\ \emph {et~al.}(2014)\citenamefont
  {Wootton}, \citenamefont {Burri}, \citenamefont {Iblisdir},\ and\
  \citenamefont {Loss}}]{wootton2014error}%
  \BibitemOpen
  \bibfield  {author} {\bibinfo {author} {\bibfnamefont {J.~R.}\ \bibnamefont
  {Wootton}}, \bibinfo {author} {\bibfnamefont {J.}~\bibnamefont {Burri}},
  \bibinfo {author} {\bibfnamefont {S.}~\bibnamefont {Iblisdir}},\ and\
  \bibinfo {author} {\bibfnamefont {D.}~\bibnamefont {Loss}},\ }\bibfield
  {title} {\bibinfo {title} {Error correction for non-abelian topological
  quantum computation},\ }\href {https://doi.org/10.1103/PhysRevX.4.011051}
  {\bibfield  {journal} {\bibinfo  {journal} {Phys. Rev. X}\ }\textbf {\bibinfo
  {volume} {4}},\ \bibinfo {pages} {011051} (\bibinfo {year}
  {2014})}\BibitemShut {NoStop}%
\bibitem [{\citenamefont {Brell}\ \emph {et~al.}(2014)\citenamefont {Brell},
  \citenamefont {Burton}, \citenamefont {Dauphinais}, \citenamefont {Flammia},\
  and\ \citenamefont {Poulin}}]{brell2014thermalization}%
  \BibitemOpen
  \bibfield  {author} {\bibinfo {author} {\bibfnamefont {C.~G.}\ \bibnamefont
  {Brell}}, \bibinfo {author} {\bibfnamefont {S.}~\bibnamefont {Burton}},
  \bibinfo {author} {\bibfnamefont {G.}~\bibnamefont {Dauphinais}}, \bibinfo
  {author} {\bibfnamefont {S.~T.}\ \bibnamefont {Flammia}},\ and\ \bibinfo
  {author} {\bibfnamefont {D.}~\bibnamefont {Poulin}},\ }\bibfield  {title}
  {\bibinfo {title} {Thermalization, error correction, and memory lifetime for
  ising anyon systems},\ }\href {https://doi.org/10.1103/PhysRevX.4.031058}
  {\bibfield  {journal} {\bibinfo  {journal} {Phys. Rev. X}\ }\textbf {\bibinfo
  {volume} {4}},\ \bibinfo {pages} {031058} (\bibinfo {year}
  {2014})}\BibitemShut {NoStop}%
\bibitem [{\citenamefont {Burton}\ \emph {et~al.}(2017)\citenamefont {Burton},
  \citenamefont {Brell},\ and\ \citenamefont {Flammia}}]{burton2017classical}%
  \BibitemOpen
  \bibfield  {author} {\bibinfo {author} {\bibfnamefont {S.}~\bibnamefont
  {Burton}}, \bibinfo {author} {\bibfnamefont {C.~G.}\ \bibnamefont {Brell}},\
  and\ \bibinfo {author} {\bibfnamefont {S.~T.}\ \bibnamefont {Flammia}},\
  }\bibfield  {title} {\bibinfo {title} {Classical simulation of quantum error
  correction in a fibonacci anyon code},\ }\href
  {https://doi.org/10.1103/PhysRevA.95.022309} {\bibfield  {journal} {\bibinfo
  {journal} {Phys. Rev. A}\ }\textbf {\bibinfo {volume} {95}},\ \bibinfo
  {pages} {022309} (\bibinfo {year} {2017})}\BibitemShut {NoStop}%
\bibitem [{\citenamefont {Schotte}\ \emph {et~al.}(2022)\citenamefont
  {Schotte}, \citenamefont {Zhu}, \citenamefont {Burgelman},\ and\
  \citenamefont {Verstraete}}]{schotte2022quantum}%
  \BibitemOpen
  \bibfield  {author} {\bibinfo {author} {\bibfnamefont {A.}~\bibnamefont
  {Schotte}}, \bibinfo {author} {\bibfnamefont {G.}~\bibnamefont {Zhu}},
  \bibinfo {author} {\bibfnamefont {L.}~\bibnamefont {Burgelman}},\ and\
  \bibinfo {author} {\bibfnamefont {F.}~\bibnamefont {Verstraete}},\ }\bibfield
   {title} {\bibinfo {title} {Quantum {{Error Correction Thresholds}} for the
  {{Universal Fibonacci Turaev-Viro Code}}},\ }\href
  {https://doi.org/10.1103/PhysRevX.12.021012} {\bibfield  {journal} {\bibinfo
  {journal} {Physical Review X}\ }\textbf {\bibinfo {volume} {12}},\ \bibinfo
  {pages} {021012} (\bibinfo {year} {2022})}\BibitemShut {NoStop}%
\bibitem [{\citenamefont {Bonesteel}\ and\ \citenamefont
  {DiVincenzo}(2012)}]{Bonesteel:2012fl}%
  \BibitemOpen
  \bibfield  {author} {\bibinfo {author} {\bibfnamefont {N.~E.}\ \bibnamefont
  {Bonesteel}}\ and\ \bibinfo {author} {\bibfnamefont {D.~P.}\ \bibnamefont
  {DiVincenzo}},\ }\bibfield  {title} {\bibinfo {title} {{Quantum circuits for
  measuring Levin-Wen operators}},\ }\href@noop {} {\bibfield  {journal}
  {\bibinfo  {journal} {Physical Review B}\ }\textbf {\bibinfo {volume} {86}},\
  \bibinfo {pages} {165113} (\bibinfo {year} {2012})}\BibitemShut {NoStop}%
\bibitem [{\citenamefont {Verresen}\ \emph {et~al.}(2021)\citenamefont
  {Verresen}, \citenamefont {Tantivasadakarn},\ and\ \citenamefont
  {Vishwanath}}]{verresen2021efficiently}%
  \BibitemOpen
  \bibfield  {author} {\bibinfo {author} {\bibfnamefont {R.}~\bibnamefont
  {Verresen}}, \bibinfo {author} {\bibfnamefont {N.}~\bibnamefont
  {Tantivasadakarn}},\ and\ \bibinfo {author} {\bibfnamefont {A.}~\bibnamefont
  {Vishwanath}},\ }\bibfield  {title} {\bibinfo {title} {Efficiently preparing
  ghz, topological and fracton states by measuring cold atoms},\ }\href@noop {}
  {\bibfield  {journal} {\bibinfo  {journal} {arXiv preprint arXiv:2112.03061}\
  } (\bibinfo {year} {2021})}\BibitemShut {NoStop}%
\bibitem [{\citenamefont {Tantivasadakarn}\ \emph {et~al.}(2022)\citenamefont
  {Tantivasadakarn}, \citenamefont {Verresen},\ and\ \citenamefont
  {Vishwanath}}]{tantivasadakarn2022shortest}%
  \BibitemOpen
  \bibfield  {author} {\bibinfo {author} {\bibfnamefont {N.}~\bibnamefont
  {Tantivasadakarn}}, \bibinfo {author} {\bibfnamefont {R.}~\bibnamefont
  {Verresen}},\ and\ \bibinfo {author} {\bibfnamefont {A.}~\bibnamefont
  {Vishwanath}},\ }\bibfield  {title} {\bibinfo {title} {The shortest route to
  non-abelian topological order on a quantum processor},\ }\href@noop {}
  {\bibfield  {journal} {\bibinfo  {journal} {arXiv preprint arXiv:2209.03964}\
  } (\bibinfo {year} {2022})}\BibitemShut {NoStop}%
\bibitem [{\citenamefont {Raussendorf}\ and\ \citenamefont
  {Harrington}(2007)}]{raussendorf2007fault}%
  \BibitemOpen
  \bibfield  {author} {\bibinfo {author} {\bibfnamefont {R.}~\bibnamefont
  {Raussendorf}}\ and\ \bibinfo {author} {\bibfnamefont {J.}~\bibnamefont
  {Harrington}},\ }\bibfield  {title} {\bibinfo {title} {Fault-tolerant quantum
  computation with high threshold in two dimensions},\ }\href@noop {}
  {\bibfield  {journal} {\bibinfo  {journal} {Physical review letters}\
  }\textbf {\bibinfo {volume} {98}},\ \bibinfo {pages} {190504} (\bibinfo
  {year} {2007})}\BibitemShut {NoStop}%
\bibitem [{\citenamefont {Fowler}\ \emph {et~al.}(2009)\citenamefont {Fowler},
  \citenamefont {Stephens},\ and\ \citenamefont
  {Groszkowski}}]{fowler2009high}%
  \BibitemOpen
  \bibfield  {author} {\bibinfo {author} {\bibfnamefont {A.~G.}\ \bibnamefont
  {Fowler}}, \bibinfo {author} {\bibfnamefont {A.~M.}\ \bibnamefont
  {Stephens}},\ and\ \bibinfo {author} {\bibfnamefont {P.}~\bibnamefont
  {Groszkowski}},\ }\bibfield  {title} {\bibinfo {title} {High-threshold
  universal quantum computation on the surface code},\ }\href@noop {}
  {\bibfield  {journal} {\bibinfo  {journal} {Physical Review A}\ }\textbf
  {\bibinfo {volume} {80}},\ \bibinfo {pages} {052312} (\bibinfo {year}
  {2009})}\BibitemShut {NoStop}%
\bibitem [{\citenamefont {Watson}\ \emph {et~al.}(2015)\citenamefont {Watson},
  \citenamefont {Anwar},\ and\ \citenamefont {Browne}}]{watson2015fast}%
  \BibitemOpen
  \bibfield  {author} {\bibinfo {author} {\bibfnamefont {F.~H.}\ \bibnamefont
  {Watson}}, \bibinfo {author} {\bibfnamefont {H.}~\bibnamefont {Anwar}},\ and\
  \bibinfo {author} {\bibfnamefont {D.~E.}\ \bibnamefont {Browne}},\ }\bibfield
   {title} {\bibinfo {title} {Fast fault-tolerant decoder for qubit and qudit
  surface codes},\ }\href@noop {} {\bibfield  {journal} {\bibinfo  {journal}
  {Physical Review A}\ }\textbf {\bibinfo {volume} {92}},\ \bibinfo {pages}
  {032309} (\bibinfo {year} {2015})}\BibitemShut {NoStop}%
\bibitem [{\citenamefont {Herold}\ \emph {et~al.}(2017)\citenamefont {Herold},
  \citenamefont {Kastoryano}, \citenamefont {Campbell},\ and\ \citenamefont
  {Eisert}}]{herold2017cellular}%
  \BibitemOpen
  \bibfield  {author} {\bibinfo {author} {\bibfnamefont {M.}~\bibnamefont
  {Herold}}, \bibinfo {author} {\bibfnamefont {M.~J.}\ \bibnamefont
  {Kastoryano}}, \bibinfo {author} {\bibfnamefont {E.~T.}\ \bibnamefont
  {Campbell}},\ and\ \bibinfo {author} {\bibfnamefont {J.}~\bibnamefont
  {Eisert}},\ }\bibfield  {title} {\bibinfo {title} {Cellular automaton
  decoders of topological quantum memories in the fault tolerant setting},\
  }\href {https://doi.org/10.1088/1367-2630/aa7099} {\bibfield  {journal}
  {\bibinfo  {journal} {New Journal of Physics}\ }\textbf {\bibinfo {volume}
  {19}},\ \bibinfo {pages} {063012} (\bibinfo {year} {2017})}\BibitemShut
  {NoStop}%
\bibitem [{\citenamefont {Dauphinais}\ and\ \citenamefont
  {Poulin}(2017)}]{dauphinais2017fault}%
  \BibitemOpen
  \bibfield  {author} {\bibinfo {author} {\bibfnamefont {G.}~\bibnamefont
  {Dauphinais}}\ and\ \bibinfo {author} {\bibfnamefont {D.}~\bibnamefont
  {Poulin}},\ }\bibfield  {title} {\bibinfo {title} {Fault-tolerant quantum
  error correction for non-abelian anyons},\ }\href
  {https://doi.org/10.1007/s00220-017-2923-9} {\bibfield  {journal} {\bibinfo
  {journal} {Comm. Math. Phys.}\ }\textbf {\bibinfo {volume} {355}},\ \bibinfo
  {pages} {519} (\bibinfo {year} {2017})}\BibitemShut {NoStop}%
\bibitem [{\citenamefont {Pfeifer}\ \emph {et~al.}(2012)\citenamefont
  {Pfeifer}, \citenamefont {Buerschaper}, \citenamefont {Trebst}, \citenamefont
  {Ludwig}, \citenamefont {Troyer},\ and\ \citenamefont
  {Vidal}}]{pfeifer2012translation}%
  \BibitemOpen
  \bibfield  {author} {\bibinfo {author} {\bibfnamefont {R.~N.}\ \bibnamefont
  {Pfeifer}}, \bibinfo {author} {\bibfnamefont {O.}~\bibnamefont
  {Buerschaper}}, \bibinfo {author} {\bibfnamefont {S.}~\bibnamefont {Trebst}},
  \bibinfo {author} {\bibfnamefont {A.~W.}\ \bibnamefont {Ludwig}}, \bibinfo
  {author} {\bibfnamefont {M.}~\bibnamefont {Troyer}},\ and\ \bibinfo {author}
  {\bibfnamefont {G.}~\bibnamefont {Vidal}},\ }\bibfield  {title} {\bibinfo
  {title} {Translation invariance, topology, and protection of criticality in
  chains of interacting anyons},\ }\href
  {https://doi.org/10.1103/PhysRevB.86.155111} {\bibfield  {journal} {\bibinfo
  {journal} {Phys. Rev. B}\ }\textbf {\bibinfo {volume} {86}},\ \bibinfo
  {pages} {155111} (\bibinfo {year} {2012})}\BibitemShut {NoStop}%
\bibitem [{\citenamefont {Herold}\ \emph {et~al.}(2015)\citenamefont {Herold},
  \citenamefont {Campbell}, \citenamefont {Eisert},\ and\ \citenamefont
  {Kastoryano}}]{herold2015cellular}%
  \BibitemOpen
  \bibfield  {author} {\bibinfo {author} {\bibfnamefont {M.}~\bibnamefont
  {Herold}}, \bibinfo {author} {\bibfnamefont {E.~T.}\ \bibnamefont
  {Campbell}}, \bibinfo {author} {\bibfnamefont {J.}~\bibnamefont {Eisert}},\
  and\ \bibinfo {author} {\bibfnamefont {M.~J.}\ \bibnamefont {Kastoryano}},\
  }\bibfield  {title} {\bibinfo {title} {Cellular-automaton decoders for
  topological quantum memories},\ }\href@noop {} {\bibfield  {journal}
  {\bibinfo  {journal} {npj Quantum information}\ }\textbf {\bibinfo {volume}
  {1}},\ \bibinfo {pages} {1} (\bibinfo {year} {2015})}\BibitemShut {NoStop}%
\bibitem [{\citenamefont {Zhu}\ \emph {et~al.}(2020{\natexlab{a}})\citenamefont
  {Zhu}, \citenamefont {Lavasani},\ and\ \citenamefont
  {Barkeshli}}]{PhysRevLett.125.050502}%
  \BibitemOpen
  \bibfield  {author} {\bibinfo {author} {\bibfnamefont {G.}~\bibnamefont
  {Zhu}}, \bibinfo {author} {\bibfnamefont {A.}~\bibnamefont {Lavasani}},\ and\
  \bibinfo {author} {\bibfnamefont {M.}~\bibnamefont {Barkeshli}},\ }\bibfield
  {title} {\bibinfo {title} {Universal logical gates on topologically encoded
  qubits via constant-depth unitary circuits},\ }\href
  {https://doi.org/10.1103/PhysRevLett.125.050502} {\bibfield  {journal}
  {\bibinfo  {journal} {Phys. Rev. Lett.}\ }\textbf {\bibinfo {volume} {125}},\
  \bibinfo {pages} {050502} (\bibinfo {year} {2020}{\natexlab{a}})}\BibitemShut
  {NoStop}%
\bibitem [{\citenamefont {Zhu}\ \emph {et~al.}(2020{\natexlab{b}})\citenamefont
  {Zhu}, \citenamefont {Lavasani},\ and\ \citenamefont
  {Barkeshli}}]{PhysRevB.102.075105}%
  \BibitemOpen
  \bibfield  {author} {\bibinfo {author} {\bibfnamefont {G.}~\bibnamefont
  {Zhu}}, \bibinfo {author} {\bibfnamefont {A.}~\bibnamefont {Lavasani}},\ and\
  \bibinfo {author} {\bibfnamefont {M.}~\bibnamefont {Barkeshli}},\ }\bibfield
  {title} {\bibinfo {title} {Instantaneous braids and {D}ehn twists in
  topologically ordered states},\ }\href
  {https://doi.org/10.1103/PhysRevB.102.075105} {\bibfield  {journal} {\bibinfo
   {journal} {Phys. Rev. B}\ }\textbf {\bibinfo {volume} {102}},\ \bibinfo
  {pages} {075105} (\bibinfo {year} {2020}{\natexlab{b}})}\BibitemShut
  {NoStop}%
\bibitem [{\citenamefont {Lavasani}\ \emph {et~al.}(2019)\citenamefont
  {Lavasani}, \citenamefont {Zhu},\ and\ \citenamefont
  {Barkeshli}}]{Lavasani2019universal}%
  \BibitemOpen
  \bibfield  {author} {\bibinfo {author} {\bibfnamefont {A.}~\bibnamefont
  {Lavasani}}, \bibinfo {author} {\bibfnamefont {G.}~\bibnamefont {Zhu}},\ and\
  \bibinfo {author} {\bibfnamefont {M.}~\bibnamefont {Barkeshli}},\ }\bibfield
  {title} {\bibinfo {title} {Universal logical gates with constant overhead:
  instantaneous dehn twists for hyperbolic quantum codes},\ }\href@noop {}
  {\bibfield  {journal} {\bibinfo  {journal} {Quantum 3, 180}\ } (\bibinfo
  {year} {2019})}\BibitemShut {NoStop}%
\bibitem [{\citenamefont {Zhu}\ \emph {et~al.}(2020{\natexlab{c}})\citenamefont
  {Zhu}, \citenamefont {Hafezi},\ and\ \citenamefont {Barkeshli}}]{Zhu:2017tr}%
  \BibitemOpen
  \bibfield  {author} {\bibinfo {author} {\bibfnamefont {G.}~\bibnamefont
  {Zhu}}, \bibinfo {author} {\bibfnamefont {M.}~\bibnamefont {Hafezi}},\ and\
  \bibinfo {author} {\bibfnamefont {M.}~\bibnamefont {Barkeshli}},\ }\bibfield
  {title} {\bibinfo {title} {Quantum origami: Transversal gates for quantum
  computation and measurement of topological order},\ }\href
  {https://doi.org/10.1103/PhysRevResearch.2.013285} {\bibfield  {journal}
  {\bibinfo  {journal} {Phys. Rev. Research}\ }\textbf {\bibinfo {volume}
  {2}},\ \bibinfo {pages} {013285} (\bibinfo {year}
  {2020}{\natexlab{c}})}\BibitemShut {NoStop}%
\bibitem [{\citenamefont {Turaev}\ and\ \citenamefont
  {Viro}(1992)}]{turaev1992state}%
  \BibitemOpen
  \bibfield  {author} {\bibinfo {author} {\bibfnamefont {V.~G.}\ \bibnamefont
  {Turaev}}\ and\ \bibinfo {author} {\bibfnamefont {O.~Y.}\ \bibnamefont
  {Viro}},\ }\bibfield  {title} {\bibinfo {title} {State sum invariants of
  3-manifolds and quantum 6j-symbols},\ }\href
  {https://doi.org/10.1016/0040-9383(92)90015-A} {\bibfield  {journal}
  {\bibinfo  {journal} {Topology}\ }\textbf {\bibinfo {volume} {31}},\ \bibinfo
  {pages} {865} (\bibinfo {year} {1992})}\BibitemShut {NoStop}%
\bibitem [{\citenamefont {Kitaev}(2006)}]{kitaev2006anyons}%
  \BibitemOpen
  \bibfield  {author} {\bibinfo {author} {\bibfnamefont {A.}~\bibnamefont
  {Kitaev}},\ }\bibfield  {title} {\bibinfo {title} {Anyons in an exactly
  solved model and beyond},\ }\href {https://doi.org/10.1016/j.aop.2005.10.005}
  {\bibfield  {journal} {\bibinfo  {journal} {Annals of Physics}\ }\textbf
  {\bibinfo {volume} {321}},\ \bibinfo {pages} {2} (\bibinfo {year} {2006})},\
  \Eprint {https://arxiv.org/abs/cond-mat/0506438} {arXiv:cond-mat/0506438}
  \BibitemShut {NoStop}%
\bibitem [{\citenamefont {Wang}(2010)}]{wang2010topological}%
  \BibitemOpen
  \bibfield  {author} {\bibinfo {author} {\bibfnamefont {Z.}~\bibnamefont
  {Wang}},\ }\href@noop {} {\emph {\bibinfo {title} {Topological {{Quantum
  Computation}}}}}\ (\bibinfo  {publisher} {{American Mathematical Soc.}},\
  \bibinfo {year} {2010})\BibitemShut {NoStop}%
\bibitem [{\citenamefont {Turaev}\ and\ \citenamefont
  {Virelizier}(2017)}]{turaev2017monoidal}%
  \BibitemOpen
  \bibfield  {author} {\bibinfo {author} {\bibfnamefont {V.}~\bibnamefont
  {Turaev}}\ and\ \bibinfo {author} {\bibfnamefont {A.}~\bibnamefont
  {Virelizier}},\ }\href {https://doi.org/10.1007/978-3-319-49834-8} {\emph
  {\bibinfo {title} {Monoidal {{Categories}} and {{Topological Field
  Theory}}}}},\ Progress in {{Mathematics}}\ (\bibinfo  {publisher}
  {{Birkh\"auser Basel}},\ \bibinfo {year} {2017})\BibitemShut {NoStop}%
\bibitem [{\citenamefont {Bazant}(2000)}]{bazant2000largest}%
  \BibitemOpen
  \bibfield  {author} {\bibinfo {author} {\bibfnamefont {M.~Z.}\ \bibnamefont
  {Bazant}},\ }\bibfield  {title} {\bibinfo {title} {Largest cluster in
  subcritical percolation},\ }\href {https://doi.org/10.1103/PhysRevE.62.1660}
  {\bibfield  {journal} {\bibinfo  {journal} {Phys. Rev. E}\ }\textbf {\bibinfo
  {volume} {62}},\ \bibinfo {pages} {1660} (\bibinfo {year}
  {2000})}\BibitemShut {NoStop}%
\bibitem [{\citenamefont {Burton}(2016)}]{burton2016short}%
  \BibitemOpen
  \bibfield  {author} {\bibinfo {author} {\bibfnamefont {S.}~\bibnamefont
  {Burton}},\ }\href@noop {} {\bibinfo {title} {A short guide to anyons and
  modular functors}} (\bibinfo {year} {2016}),\ \Eprint
  {https://arxiv.org/abs/1610.05384} {arxiv:1610.05384 [quant-ph]} \BibitemShut
  {NoStop}%
\end{thebibliography}%

\end{document}